\documentclass[runningheads]{llncs}
\usepackage{ifpdf}
 \ifpdf
\usepackage[pdftex]{graphicx}
\pdfoutput=1
\newcommand{\mypix}{jpg}
 \else
\usepackage[dvips]{graphicx}
\newcommand{\mypix}{bmp}
 \fi

\newcommand{\colo}[1]{}

\usepackage{epsfig}
\usepackage{subfigure}
\usepackage{calc}
\usepackage{amssymb}
\usepackage{amsmath}

\def\F{\mbox{\rm{I}\hspace{-.20em}\rm{F}}}

\usepackage{url}
\spnewtheorem{defi}{Definition}[subsection]{\bfseries}{\rmfamily}
\spnewtheorem{prop}{Proposition}[section]{\bfseries}{\rmfamily}
\spnewtheorem{coro}{Corollary}[section]{\bfseries}{\rmfamily}
\spnewtheorem{theor}[subsection]{Theorem}{\bfseries}{\rmfamily}

\makeatletter
\def\url@chrisstyle{%
  \@ifundefined{selectfont}{\def\UrlFont{\sf\scriptsize}}{\def\UrlFont{\ttfamily\sf\scriptsize}}}
\makeatother
\usepackage[dvipsnames]{xcolor}  
\definecolor{OliveGreen}{rgb}{0,0.6,0} 

\usepackage{fancyhdr}                           
\fancypagestyle{firstpage} 
{
   \fancyhf{}
   \fancyfoot[C]{\vspace{-12mm}\thepage}

   \fancyfoot[L]{\vspace{-5mm}\noindent\rule{4.8in}{0.4pt}
     \scriptsize{
         \textcopyright \hspace{0.5mm} Christopher D. Clack and Nicolas T. Courtois 2018-2019 \\
       This work is licensed under a {Creative Commons Attribution 4.0 International License (CC  BY):}  
       \url{https://creativecommons.org/licenses/by/4.0/} \\
       Provided you adhere to the CC BY license, including as to attribution, you are
       free to copy and redistribute this work in any medium or format and remix, transform,
       and build upon the work for any purpose, even commercially.                  \\
       \vspace{1mm}
     }
 }
}

\begin{document}

\def\mytitle{Distributed Ledger Privacy: Ring Signatures, M\"{o}bius and CryptoNote}
\titlerunning{DLT Privacy: Ring Signatures: M\"{o}bius \& CryptoNote}

\title{\Large \mytitle}

%

\author{
Christopher D. Clack 
\and
Nicolas T. Courtois 
}

\institute{
Centre for Blockchain Technologies\\
Department of Computer Science, University College London}

\maketitle

\thispagestyle{firstpage} 

\abstract{
%
%
Distributed ledger and blockchain systems are expected to make financial systems easier to audit,
reduce counter-party risk and transfer assets seamlessly.
The key concept is a
token controlled by a cryptographic private key for spending,
and represented by a public key for receiving and audit purposes.
Ownership transfers are authorized with digital signatures and recorded on a ledger visible to numerous participants.
Several ways to enhance the privacy of such ledgers have been proposed.
In this paper we study two major techniques to enhance privacy of token transfers
with the help of improved cryptography:
M\"{o}bius \cite{Mobius} and CryptoNote \cite{CryptoNote20}.
The comparison is 
illuminating: 
both techniques use ``ring signatures''
and some form of ``stealth addressing'' or key derivation techniques,
yet each does it in a completely different way.
M\"{o}bius is more recent and operates in a more co-operative way (with permission) and
is not yet specified at a sufficiently detailed level. 
Our primary goal is to explore the suitability of these two techniques
for improving the privacy of payments on cryptographic ledgers.
We explain various conflicting requirements and strategic choices
which arise when trying to conceal the identity of participants and the exact details of transactions in our context
while simultaneously enabling fast final settlement of tokens with a reasonable level of 
liquidity. We show that in these systems,
third-party observers see obfuscated settlement.
We finish with a summary of explicit warnings and advice for implementors of such systems.


%
%
%


\colo{black}

\newpage
\section{Introduction}
\label{sec:GenIntro}

Money and other assets have been electronic for decades,
and it appears paradoxical
that we have
very fast (``low-latency'') trading and
yet very slow centralized ownership and final settlement of assets, including cash.
Distributed ledgers (DLs) are an emerging technology that attempts to solve this problem,
and there is the potential for this to create a major paradigm shift
in how financial
assets can be stored and transferred.
Ownership and control is promised to become more decentralized and
a new type of distributed and cooperative financial infrastructure is being planned.
%
%

In Nakamoto-type \cite{SatoshiPaper} systems using cryptographic tokens (``crypto tokens''),
these tokens represent coins and are attributed to pseudonymous entities
(typically based on cryptographic public keys).
Transfers of ownership are performed with digital signatures and
recorded on a ledger composed of blocks visible to numerous participants.
High visibility, together with the imperfect nature of pseudonymity, leads to a loss of privacy.

In this paper we explore some of the cryptographic technology that is being used to provide
improved privacy for blockchains, and in particular we study two systems in greater depth: CryptoNote \cite{CryptoNote20}
and M\"{o}bius \cite{Mobius}.  These systems are viewed as two fundamentally different methods which use both
ring signatures and some form of stealth addressing or key derivation technique
in order to enhance the privacy of asset transfers.
In spite of some similarities we show that there are fundamental differences in how
the two systems use public keys and the identities of transaction participants.
Other strong differentiators are co-operative operation
versus the ability to enroll extra participants without their permission,
questions of how they need to rely on additional authentication and communication channels instead of committing storage
to the blockchain, how they obfuscate the current state of the ledger and the 
transaction flow, and how future actions can compromise privacy of previous transactions.
In contrast to the CryptoNote protocol that is used in at least two major cryptographic currencies (``crypto currencies'') \cite{bytecoin,monero}, M\"{o}bius
is a more recent, substantially simpler, solution
that is however not yet sufficiently specified for practical implementation.


We explore the existence of fundamental trade-offs
between settlement, liquidity and privacy (which includes
the privacy of the actual identities that are linked to pseudonymous
cryptograms,  the privacy of the details of individual transactions,
and privacy in terms of linkability of separate transactions).
Our goal is to improve the understanding
of what such systems can or cannot do in order to
build fast yet private payment and settlement systems.

This paper is written for a broad audience including blockchain and cryptographic engineers,
financial systems engineers and financial services practitioners.
Although appropriate technical terms and vocabulary are used, mathematical and cryptographic discussions
stay at a high (less detailed) level and in this way the paper aims to enable an increased understanding of
modern financial-cryptographic technology by a broad range of stakeholders.
We hope that the issues and views raised in this paper will stimulate debate
and we look forward to receiving feedback.

\subsection{Outline}
\label{sec:IntroOrganized}

This paper explores the interaction between the two worlds of cryptogrphic assets and
privacy in financial markets.
Cryptographic tokens represent a certain type of new technology push,
and the question arises as to whether these new systems are {\bf at all} fit
for purpose in a practical commercial context of financial technology.
This question is hard to answer because distributed ledgers are a new and highly disruptive technology,
and because issues of privacy  and especially cryptographic privacy (which is the main area that requires improvement)
are traditionally very difficult and poorly understood.

The paper is organized as follows:
Section~\ref{sec:WhyPrivacyDifficult} provides an overview of the emerging ecosystem of cryptographic assets and blockchain privacy;
Section~\ref{sec:cryptomethods} explores some of the cryptographic technology that is being used to provide improved privacy for blockchains;
Section~\ref{sec:MobiusvsMonero} investigates two systems in greater depth (CryptoNote and M\"{o}bius);
Section~\ref{sec:discussion} provides a broader discussion, facilitated by the detail covered in the previous sections --- including possible improvements to M\"{o}bius, and
some advice to implementors (summarised in two tables titled: {\em ``Beware Crypto!''}); and
Section~\ref{sec:conclusion} concludes.

\newpage

\section{Crypto Assets and Blockchain Privacy}
\label{sec:WhyPrivacyDifficult}



\label{sec:TokenPrivacyProblemIntro}
A new form of digital economy has emerged in recent years:
an economy of tokens stored as entries on a {\em distributed} or {\em decentralized} ledger (``DL'') \cite{swanson2015}
where ownership is primarily controlled through cryptographic private keys.  These cryptographically-controlled tokens
(``crypto tokens'') may be representative of real or abstract quantities, and may be created, awarded, stored and exchanged \cite{tokens}.
\label{sec:TokenPrivacyProblemDigitalAssets}

Distributed ledgers have more functionality,
and therefore more value than,
merely the possibility to transfer tokens and exchange them against goods
or fiat currency.
For example, they can create new forms
of control over a variety of assets. 
They are able to serve a variety of purposes
such as shared ownership or managing various rights in a distributed
software system or in a business ecosystem.
They can convey various combinations of privileges such as
voting, execution of code or scripts by one or many participants,
participation in auctions,
or the control of write or read access rights to data.

In this new form of digital economy transactions typically obey integrity rules (for example, a rule that a single token
can only be transferred to one other owner --- it cannot be ``spent twice'' by transferring that single token to two
new owners) and these rules may be enforced by a combination of system software and/or privileged individual or collective entities
(such as ``miners'' \cite{vranken}) with discretionary powers to decide which transactions are eventually accepted.
%


In a ``pure'' crypto-token economy, the tokens themselves are valued assets (rather than being representative of
some physical asset),
and for practical purposes ``ownership'' is the ability to exercise control of these crypto assets on the DL --- e.g. to transfer control
to another entity. Pure crypto assets and tokens that represent physical assets may coexist on the same DL, and both will be
subject to the force of law in the appropriate jurisdiction (though determining the applicable law may
be problematic for a DL with nodes in more than one jurisdiction).

The records of the transfer of ownership of crypto tokens are typically held in a cryptographically-secured ``block'', and
each new block containing new records is cryptographically attached to the
previous block.  A ``blockchain'' is therefore a sequence of these connected blocks.  The way that the addition of a new block is agreed and added to the blockchain,
and the finality of that addition, varies between systems.  If a block, once added, can never be removed, and if the chain is strictly linear and
cannot ``fork'' \cite{muller},\footnote{Also see: \url{https://en.wikipedia.org/wiki/Fork_(blockchain)}}
then once the change of ownership of a token has been agreed and committed to a block on a blockchain, the record
of new ownership is immutable except by the action of a new block added to the blockchain to record a further transfer of ownership.  Thus, addition
of the block to the chain represents finality of the transfer of ownership within the DL: but if the token represents some physical asset then finality of transfer on
the chain may not align
with transfer of ownership of the real-world asset.

\subsection{Crypto Currencies, Native vs. Added Functionality}


One of the major applications of crypto tokens is to use them for payment and as a currency.
Such systems are often called ``crypto currencies''\footnote{Or ``cryptocurrencies''.}
--- for example Bitcoin, Monero and ZCash (also known as ZeroCash).
The Bank for International Settlements views crypto currencies as a subset of ``digital currencies'',
where the latter would include any currency in dematerialised form \cite{bis}.
Traditional government-issued currencies and crypto currencies alike can serve as
as a medium of exchange, as a unit of account, and as a store of value.
It is hard however to conceal the fact that they are very far from being similar 
and natively do not achieve these three functions equally well. 
For example most crypto currencies are traded against each other and against
traditional currencies such as the US Dollar or Euro,
and may be used to pay for goods or services and some merchants actually accept them.
However the practicality of using crypto currencies as a store of value is contentious because of the
highly variable exchange rate against traditional fiat currencies \cite{fsb}.

Most fundamental differences here stem from what is native, and what needs to be added.
The digital world is natively characterized by high speed, high efficiency, poor security and poor privacy.
In contrast, the physical world is natively characterised by slow speed, poor efficiency, better
security (though protection against forgery has required time to develop) and better privacy.
The basic level of trust between users is low in both worlds, but in the physical world low trust
between users has been replaced by high trust in central agencies (e.g. banks), regulators, and
the courts of law.  Due to the operation of regulation and law, privacy is quite strong in the physical world,
but requires a lot of attention in the digital world.

Due to the inefficiencies and risk involved in transferring cash in the physical world, it is mostly
avoided and instead the "transfer" of money is achieved by changing account balances at each
end of the transaction.  Thus, for large transactions between financial institutions there has already
been a de-facto dematerialisation (or digitisation) of cash, and this dematerialisation depends on
the trust relationships between the retail, commercial and central banks, and with a large number
of financial intermediaries.
Dematerialisation of money is not the same as the creation of a crypto currency.
In most crypto currencies there is a return to the notion of a currency artefact (a ``coin'') rather than
relying on the modification of account balances at each end of a transaction.
In a crypto currency  the coin is however a digital artefact, and issues of trust are established in a new way
--- rather than trusting in strongly-regulated intermediaries and the courts of law,
users place their trust in cryptography
and in our ability to secure computer systems against attacks.
Thus, crypto currencies are typically designed to be {\em disintermediated}
and {\em decentralised} with no requirement
for regulated banks or other financial intermediaries.

Different crypto currencies may differ in their implementation.  The most well-known is Bitcoin, which provides
a transparent ledger with cryptographic hashes of public keys serving as pseudonyms (thus, all participants can see
all transactions, but identities are given as numbers derived from cryptographic keys, not as the names of people or organisations);
Bitcoin also
provides a certain (limited) scripting capability, and
uses ``miners'' 
who are able to create new coins and approve transactions.

For pure crypto tokens, finality of transfer on the blockchain implies that once a transfer is recorded on the chain the receiving party should no longer
be exposed to the risk of the sender defaulting (counterparty risk).  However, this also depends on both the contractual agreement between the blockchain participants
and the governing law.  The receiver also remains exposed to the risk that the DL itself may suffer technology failure,
cybersecurity events (theft, crypto bugs), or financial collapse of the infrastructure provider that is running the DLT.  Furthermore, if there is an intention
to trade the crypto asset against a real-world asset such as a fiat currency, then a store of crypto assets will be subject to market risk because
of the highly volatile exchange rates. 

\subsubsection{Inter-bank payments.}

When 
banks make domestic payments between themselves, whether on behalf of their clients or on 
their own behalf, and whether or not they are linked with some other transfer,\footnote{For example, when settling an equities trade, 
the transfer of ownership of the equities will be linked to, and occur 
simultaneously with, the payment for those equities.} those payments may be conducted directly between two banks or via a 
central bank, and may also involve an intermediary 
bank.  For large banks it is unlikely that domestic payments would involve more than one intermediary, whereas 
cross-border payments may involve several.\footnote{We will ignore the mechanics of deferred-net and real-time-gross settlement of payments, except to note in passing the 
possible need for ``liquidity-saving mechanisms''  \cite{martin} when using crypto currencies for inter-bank payments.}

%

Payments may be effected 
between banks not as transfers of physical money but 
as changes in the recorded debt obligations between two banks (in ``correspondent'' banking each bank holds an
account with the other, and the recorded account balances may be modified to reflect a payment), 
but if these debt obligations become significantly unbalanced a real transfer of value is likely to be required. 

The difference between a recorded debt obligation (``Bank A agrees that it owes me \pounds Xm'') and actual value 
(``Bank A has transferred to me the ownership, with finality, of a financial instrument worth \pounds Xm'') 
becomes important in the case that Bank A becomes unable to pay its debts, or suffers technical problems 
that delay Bank A's ability to turn its debt obligation to another bank into actual value for that other bank.  
When a institution holds money for a client it has a debt obligation to the client (it must pay the money back on demand or according to
the terms of a commercial agreement), and if it is unable to pay the debt (or any due interest on the debt) when due this is 
known as ``default'' (credit risk).

Fortunately, central banks are notable in that their risk of default (with a few notable historical exceptions) is very much lower than that of commercial or retail banks. 
This difference in risk leads to the preference to settle payments in ``central bank money'' (i.e. either cash issued by a central bank, or an account balance held with a central bank) 
rather than ``commercial bank money'' (i.e. an account balance held with a commercial bank).  Central bank money functions well as a medium of exchange, with fast real-time settlement available, and as a unit of account, but functions less well as a store of wealth due to fluctuations in value (explored further below).\footnote{Central banks also have other roles such as printing money (e.g. cash) and managing the economy (e.g. by controlling the amount of cash in circulation).}
Commercial bank money also functions well as a medium of exchange, with fast settlement (slower if deferred settlement is used), but in terms of a unit of account or a store of value, commercial bank money may be subject to both (i) the risks associated with the underlying sovereign (central bank) currency and (ii) the risks of commercial bank default (explored further below).

Typically a crypto currency coin will be an asset that can be held and potentially stolen or lost or destroyed.  
A crypto currency might be designed to incorporate a commercial obligation from the network to pay the user's crypto coin account 
balance back to the user on request,\footnote{Perhaps as a sovereign currency according to an agreed exchange rate.}  
and if it does {\em not} do so when due, this would be a ``default''.
In other cases however there will be no such guarantee and therefore no ``default''.  

Some crypto currencies are being used as a medium of exchange, though their use is not widespread.\footnote{A list of some shops that accept Bitcoins, compiled in September 2018, is provided at \url{https://99bitcoins.com/who-accepts-bitcoins-payment-companies-stores-take-bitcoins/}}
Crypto currency transaction setlement might eventually be extremely fast, but currently is slow and does not perform well in terms of bandwidth.\footnote{\cite{bisannualreport} reports that Visa processes 3,526 transactions per second, whereas  Bitcoin processes only 3.3 transactions per second (2017 figures). Crypto currency bandwidth is improving, but remains slow compared with Visa, Mastercard and PayPal.}  Crypto currencies do have the potential to reduce the risks associated with failure of intermediaries, but this improvement must be balanced against privacy concerns related to increased disclosure. Crypto currencies vary in their design and this will affect their functionality as a unit of account or a store of wealth.

The risks inherent in different kinds of currency include the following:
\begin{description}
\item[Market risk.]
All currencies suffer market risk in that their exchange rate with other currencies varies with time.  
Here
we also
include liquidity risk: the possibility that demand for a currency becomes 
very limited, perhaps even zero, and that this might persist temporarily or permanently.
Some central bank (sovereign) 
currencies have more stable exchange rates than others.  
Commercial bank money is typically denominated in a sovereign currency and so inherits
the market risk of the denominated currency.  
Crypto currencies might include a guaranteed exchange rate to a single sovereign currency, or perhaps to a basket of foreign currencies,
in which case they inherit the market risk of that sovereign currency (or those sovereign currencies).
Crypto currencies that do not have guaranteed exchange rates 
often have
large exchange rate volatility and very high market risk.  
\item[Risk of default (credit risk).]
As explained above, central banks have very low credit risk, commercial banks have medium credit risk, and crypto currency networks either have  high credit risk (e.g. if they offer a guaranteed exchange rate), or are structured so that they do not have debt obligations (e.g. with no guarantees) 
or additional mechanisms are employed to reduce credit risk.\footnote{Discussion of such mechanisms is outside the scope of this paper.}
\item[Risk of collapse.]
Central bank money, commercial bank money, and crypto currency money, are all liable to the risk of collapse.  
This can occur as a result of many factors and with both central bank money and commercial bank money it is likely to be
immediately linked to default.  With all currencies, if liquidity drops to zero then collapse may be imminent.  Similarly, adverse
exchange rates may cause a currency to lose substantial value and collapse may soon follow.  With crypto currency money,
the network itself might collapse either by the loss of all users or by some technological fault in the system (for example, 
causing all account balances to be lost and unretrievable, or inability to process transactions).  Crypto currencies are particularly at 
risk with respect to the immaturity of their technology base.
\item[Risk of inflation, devaluation and demonetisation.]
The value of any currency will vary according to other factors that may for example be linked to the amount of that currency in 
circulation and how frequently the currency is used.  
The value of central bank money often reduces over time due to inflation, and its value in relation to other currencies fluctuates (market risk) and may be forcibly devalued. Holders of cash (e.g. individuals and retail businesses) may also be affected by demonitisation policies: for example, see \url{https://en.wikipedia.org/wiki/2016_Indian_banknote_demonetisation}.  With crypto currencies, similar dynamics may
apply in different ways according to the design decisions for the crypto currency (e.g. fixed versus variable money supply).
\end{description}

\subsection{Distributed Ledger Technology in Financial Markets}


As both regulation and competition have increased over the past decade there
has been a drive towards standardisation (and commoditisation)
in the financial markets.
DLT is attractive to financial institutions for several reasons:
\begin{itemize}
\item
primarily as mechanism for reducing infrastructure costs (including
reduction of compliance costs where DLT provides automation of compliance reporting);
\item
to reduce intermediaries (such as correspondent banks) and therefore provide faster, less risky and potentially more private payments (though settlement issues need to be addressed);
\item
to reduce transaction costs;
\item
to improve cash management and utilisation (via just-in-time payments);
\item
for improved interaction with the broader DLT ecosystem (e.g. where assets are held on DLT, it may be easier if payments
for those assets are also processed via DLT); and
\item
to stimulate innovation in new products, services, and workflows.
\end{itemize}

However, there are some potential concerns for the use of DLT in financial markets.  For example, when a bank currently interfaces with a centralised payments system, some key regulatory aspects of multilateral payments processing are the responsibility of the central system,
whereas if a bank were to process multilateral payments itself (in effect, becoming a designated payments system), then the regulatory burden on the bank might increase in nature or scale.
It is not clear how regulatory requirements to destroy data would be managed,
nor how that could differ between participants (for example, if one participant were subject to a retention hold)?

A financial institution might in future be a participant in many DLs, each of which may or may not
use a blockchain.  For each such DL, the financial institution must choose an implementation mechanism, which could for example be:
\begin{itemize}
\item
to host the DL node itself;
\item
to use a separate infrastructure company to run the DL node (perhaps owned by and acting as an agent for the financial institution that it serves) ---
this could be preferred by banks that wish to segregate risk for operational or regulatory reasons;\footnote{For example, a bank might be uncomfortable with processing (validating) payments for other banks.}  or
\item
to use an API to connect to a Financial Market Infrastructure\footnote{See \url{https://www.bankofengland.co.uk/financial-stability/financial-market-infrastructure-supervision}}
company (FMI) which will run the node
on its behalf (and perhaps also run the nodes for several other participants).\footnote{It is unlikely that financial institutions will all embrace the new technology at the same
rate, and it is likely that for a while there will be a mixture of modes of operation.}
\end{itemize}


\subsection{The Importance of Blockchain Privacy}

The section starts with two core observations before exploring privacy in financial markets and for inter-bank payments.  The two
core observations (further explained below) are:

\vspace{6pt}

{\em
Observation 1: Privacy is a fundamental requirement for crypto assets.

\vspace{6pt}

Observation 2: Privacy is a technological requirement for scalability.
}

\subsubsection{Privacy as a Fundamental Requirement.}
The advent of crypto assets where asset ownership and transactions are recorded and enacted on
a DL, with or without a blockchain, brings a change in risk profile.  A key change for public,
unpermissioned, DLs is that owners of
digital assets may become more directly exposed to criminal activity (e.g. assets are no longer protected inside a
secure perimeter inside a bank).
Even with private, permissioned,
DLs the owners of digital assets are more directly exposed to
the inquisitiveness of  other participants and to external criminal actions
if the private DL is successfully breached.
Furthermore, in highly regulated areas of business the disclosure of certain types of data is illegal
(and recent European GDPR legislation\footnote{See \url{https://eur-lex.europa.eu/eli/reg/2016/679/oj}}
has introduced further widespread legal controls relating to data privacy).
An obvious response to these increased risks is to require yet greater 
privacy and security:
privacy of asset holdings, privacy of all transaction details (e.g. sender identity, recipient identity,
amount transferred, and time of transfer), and security against fraud and theft.

Cryptography provides a partial solution, though not devoid of serious problems.
Owning a crypto asset via a private cryptographic key (which is simply a sequence of bits) is fragile and potentially dangerous.
If such a sequence of bits were to be stolen, the thief could steal the assets controlled by the sequence of bits
and there may be no way to recover those assets.
Alternatively, a person with malicious intent could alter the operation of electronic systems and cryptographic protocols, perhaps causing damage that requires repair, perhaps rendering the cryptographic keys useless, or interfering with the ownership of assets and records of whether transactions are approved or completed.  This might make some assets unrecoverable.
Wealthy and/or politically exposed users might be particular targets for criminal activity if their asset holdings and transaction are known.
Privacy in high demand\footnote{If privacy is not mandated by regulators,
it is likely that commercial competition will lead to the development of financial services
which offer enhanced privacy, in order precisely to attract these wealthy costumers.}.
For all the reasons above, it is worth considering whether there is perhaps a {\em fundamental requirement} to provide a high level of privacy in DL and blockchain systems.
In Sections~\ref{sec:cryptomagic} and \ref{CryptoImplementationPitfalls} we illustrate some of the many pitfalls of cryptographic security and its implementation.

\subsubsection{Privacy as a Technological Requirement.}
\label{TokenPrivacyProblem:as:BlockchainSpaceEconomy}

Most existing distributed ledger systems duplicate storage and computations tremendously:
many different entities do (or check) the same computations again and again,
and this is wasteful.
Duplication and decentralization is also undertaken in order to inform market participants about activity in the markets,
which can be imposed by legislators and regulators \cite{FasterBitcoin}.
Furthermore, a blockchain grows monotonically, with the entire chain being replicated on all nodes.\footnote{Increased usage of computer memory also occurs as a result of slow processing.
For example, for about one month
in 2017, just before the BitcoinCash fork, 
there was a permanent backlog of up to 200,000 bitcoin transactions
which the bitcoin network was unable to ``publish'' or approve (see \url{http://blog.bettercrypto.com/?p=3510}).
Interestingly this backlog was resolved shortly after the split.}
Accordingly, 
``space'' (computer memory) in blockchains is a precious and therefore expensive resource.
From this point of view many blockchain privacy techniques can be seen as imposed
by the necessity to decrease the disclosures due simply to the high price of
``publishing'' entries in DLs.
Privacy appears to be a {\em basic technological requirement} for efficient operation of large-scale, long-lived DLs.

\subsubsection{Privacy in Financial Markets.}
\label{sec:TokenPrivacyProblemIntro:TokenPrivacyProblem}

Many financial markets for trading assets are ``lit'' markets where information about orders and trades are disclosed
to market participants.  However, not all information is disclosed; typically (and importantly) the identities of those who have submitted
orders are not disclosed, and the identities of the counterparties to
a trade are not publicly disclosed.\footnote{However, a large number of intermediaries know key information: for example, brokers,
custodians, clearing houses, settlement banks, and potentially several correspondent banks involved during the processes
of settlement and payment.}
%
When assets are not traded, privacy is even better and asset holdings are not disclosed.
Traditionally, privacy relating to ownership of assets and details of transactions has been a matter of trust in and between highly-regulated
institutions.

By contrast, modern peer-to-peer systems such as distributed ledgers
have the ability to dispense with such trusted third parties.
However, trust is problematic when dealing directly counterparties,
and especially with multiple pseudonymous counterparties.
Therefore many DLT platforms attempt to establish trust
via transparency. Many implementations of DL and blockchain technology {\em increase}
disclosure and {\em reduce} privacy.

The identities of counterparties to a transfer of ownership are often disclosed,
and asset holdings may also be disclosed (or may be inferred).
The fact that these identities are obfuscated by pseudonyms, and further via cryptography
does not mean that they cannot be derived by other means, and once the true identities behind the cryptographic keys
are known, comprehensive information becomes readily available.
Example information that may be disclosed includes:


\begin{itemize}
  \item Some identifiers, e.g. pseudonyms of participants.
  \item Amounts and types of assets being transferred.
  \item Account balances and various statistics about their age, probability distribution, velocity, mobility etc.
  \item Timing and linking of transactions.
\end{itemize}

This leads to important data leaks during operation with progressive erosion of privacy via future events.
The goal of many improved DL/blockchain solutions is to mitigate
rather than completely remove these disclosures
(which may be impossible or require a fundamental redesign of the entire DL/blockchain system).
Unfortunately cryptographic technology can not solve all practical problems equally well.
Cryptography makes unbreakable locks in areas of encryption,\footnote{With decades of maturity since the Second World War, encryption technology is extremely strong --- when used --- in contrast almost nothing is ever encrypted in blockchain systems.} and it can do integrity and authenticity extremely well for business purposes.\footnote{Digital signatures do an extremely good job of ensuring the integrity of transactions.}
However cryptography is weak when trying to achieve more difficult objectives,  and struggles to solve privacy problems in a definitive and clear-cut way;
it will {\em obfuscate} instead of concealing completely.

\subsubsection{Privacy for Inter-bank Payments.}

Current systems for inter-bank payments do not provide complete privacy: domestic payments may pass through FMI systems,
intermediary (correspondent) banks, and the central bank.  The current systems are also not entirely
risk-free (for example, the use of a correspondent bank introduces exposure to the risk of that bank failing).  And revealing data to a correspondent
bank can be of more concern than revealing data to an FMI, despite regulatory control, due to the aggressive
competition between many banks.

There are many drivers of privacy concerns for payments between large financial institutions. For example, there are regulatory concerns regarding
the confidentiality of certain types of data (and requirements to delete certain data), and participation in a DL would provide more
direct exposure of that data (both because of the transparency requirements of the consensus mechanism and because the protection of
data cannot be entirely controlled by each bank if it is replicated at other bank nodes).
There are also commercial and operational drivers:

\paragraph{Commercial drivers of privacy concerns.}
The requirement to retain privacy of information such as transaction amounts and trading strategies is often driven by competitive imperatives,
such as the desire to reduce predatory action by other banks.\footnote{Privacy from other banks is often viewed as more important than privacy from FMIs, who are often assumed not to have predatory intent.} For example, during a systemic event such as a credit crunch, payments might be controlled due to cash liquidity issues and a particularly constrained bank might be detected from network activity.  In general, analysing patterns in large numbers of payments over time (together with other sources of information)\footnote{A bank's payments reflect a mix of its own business and that of its clients; some information about a bank may be derivable from network activity but not necessarily full information. However, where other information is available that can be correlated with payments information then it may become possible to infer a bank's business separately from that of its clients.} may detect a change in behaviour indicating an exploitable vulnerability that could be incrementally explored,
for example by delaying payments, applying commercial pressure during contract negotiations,
or reducing the market value of assets the bank is known to hold and may need to sell.

\paragraph{Operational drivers of privacy concerns.}
The migration from legacy payments infrastructure to a DL/blockchain infrastructure raises
serious new concerns relating to privacy and security.
With traditional infrastructure, the data and processes to be protected are either fully controlled by the financial institution or by a highly-regulated FMI.
By contrast, with a DL/blockchain architecture the data and processes will be replicated across a large number of participating financial institutions.
The immediate concerns relating to a DL/blockchain architecture are:

\vspace{6pt}

\begin{enumerate}
\item
The weakest link risk:
that security/privacy of a bank's data and processes with respect to external attackers
will depend on the security and privacy afforded by the least secure participating institution.
\item
The lack of privacy risk: that sensitive data and processes will be replicated across all participating institutions
and may become known to all institutions
\item
The obfuscation risk: that ``privacy via cryptographic obfuscation'' (less than via strong encryption)
will not provide sufficient privacy to match banking requirements, which are more stringent than for many other industries.
\item
The operational scale risk: that the amounts of data involved
(e.g. in terms of holding transaction data for all other participating institutions, on an ever-increasing blockchain, forever)
and their operational integrity might be prohibitively expensive to manage.
The processing requirements and required system availability and security
might be similar to providing the operational availability of CLS functionality at each node. 
Blockchain compression will therefore be an important consideration,
and one way to achieve blockchain compression is increased privacy (less shared data on the blockchain).
\item
The validation burden risk: that there might be additional burdens involved in holding/processing data for a validation-only purpose (i.e. not as a counterparty to a trade) for other institutions:
\begin{itemize}
\item
there might be an additional regulatory burden (due to becoming a regulated payments infrastructure provider);
\item
there might be an additional staff oversight burden;
\item
a new privacy policy would be necessary for other data which is now replicated on the institution's own node (including protection/retention/deletion), and a new data processing policy would be necessary for the institution's validation activities;
\item
it is currently unknown whether there might be a regulatory difference between ``seeing'' data versus ``holding a copy'' of data,
and whether the encryption of this data would relax the regulatory obligations;
\item
there might be a ``guilty knowledge'' risk (i.e. the holding of data for other participants might imply responsibility to check and potentially act on that data, so ignoring the data is risky);
\item
there will doubtless be a network agreement such that each participant will have a responsibility to the whole DL network, and there might be a financial or reputational risk if a node were temporarily unable to engage\footnote{If a large financial institution's node were to fail, it would take a prohibitively long time to recreate the node by copying data from another node, and therefore every node would keep its own full backup of all data and it is likely that this would be configured with a ``shadow'' node running in another location ready to take over immediately following failure of the primary node (i.e. each institution would operate two nodes).
Thus one of the advertised benefits of DL/blockchain systems --- that all data is automatically backed-up in multiple locations --- would not hold in this use case.}
or if it were compromised by hackers.
\end{itemize}
\end{enumerate}

Currently, a precautionary principle is used in that data is only shared with those that ``need to know''.
Although compliance reports must be made to regulators, the need for privacy in financial transactions has been highlighted by the
Bank of England's privacy objective which is to explore ``how DLT based systems could be configured to ensure that no party (except for the regulator)
was able to infer details about transactions which they were not counterparty to, including ensuring that participants in the consensus process did not have full visibility of transaction details''.\footnote{\url{https://www.bankofengland.co.uk/-/media/boe/files/fintech/chain.pdf}}

\subsection{Difficult Choices for Blockchain Privacy}


Implementing blockchain privacy is not straightforward, since transparency (contrary to privacy)
is one way to establish trust where the participants do not trust each other (which might hold regardless of whether the
system is permissioned or permissionless).\footnote{For permissioned systems where the participants are
non-competitive and fully trust each other, privacy is easier to establish.}
Yet privacy remains in high demand when the
participants do not trust each other.
Here we discuss some of the issues that arise when attempting to introduce privacy into a system
where trust is problematic.

\subsubsection{Privacy.}
\label{subsec:blockchain:integrity2}

Specific aspects of privacy in a blockchain context include the following:

\begin{itemize}
  \item Are the main/fundamental (as opposed to ephemeral) identities of participants protected? 
  \item Can different transactions by the same sender be linked? 
  \item Can different transactions sent to the same recipient be linked? 
  \item Is it possible for an attacker to exploit timing side channels (timing and amounts of transactions)? 
  \item Is it possible for an attacker (or anyone other than the counterparties to a transaction) to see the amounts and timing of a transaction?
  \item Are the holdings of crypto tokens protected
  against theft and are larger holders protected from unnecessary exposure?
  \item When transactions are validated, who does the validation and how much data do they see?
  \item Are transaction histories protected?
\end{itemize}

Where protection is claimed, it is important to consider {\em how} such protection is provided, and how that protection could potentially
be attacked and subverted.  It is also important to understand the difference between different strengths of cryptographic encryption and
the limitations of cryptographic obfuscation.

\subsubsection{Privacy and Integrity.}

\label{subsec:blockchain:integrity}
Regardless of issues of trust between participants, all participants must be able to trust the integrity of the technology platform.
The main integrity questions to consider, and some implications for privacy, are:

\begin{itemize}
  \item Is ownership of crypto tokens secure so that each crypto token always has a legitimate owner and the legitimate owner (and only the legitimate owner) can
  control that token (e.g. for a crypto currency, each coin has an owner, and only the owner
  can spend the coins and in doing so transfers ownership to the recipient)?
  \item
Furthermore can this be ensured and audited
without revealing the ownership of all tokens
and with the transaction current state and past history remaining obfuscated?
  \item Are withdrawals and spending operations unique and can accidental or intentional double spending be detected? How can this be achieved without
  disclosing any details of the transaction to other participants? 
  \item Do the inputs and outputs of each transaction match, so that crypto tokens can neither be created from nothing nor destroyed? How can this be achieved without disclosing the details of the transaction?
  \item Does the system incorporate additional audit mechanisms (e.g. so that a regulator can view the details of all or selected transactions), and if so
  how is this achieved without further compromising the privacy of participants?  
  \item Does the system ensure that operations are not vulnerable to misbehaviour of particular network participants?
  For example, is the blockchain consensus algorithm secure against abuse by powerful entities?  
  \item
  If fraud occurs can it be detected rapidly and remedied before any harm is done?
  \item
  Can it be remedied in real-time without imposing a burden on innocent users?
  \item If the system requires additional off-chain channels to be used, how (if at all) does the system support the authentication of those channels?
  Does the use of off-chain channels improve privacy, and if so how?  
\end{itemize}

For each of the above questions, it is also important to establish the details of how these aspects are assured, including all
assumptions and premises, in order to determine the degree to which the technology platform can be trusted
(and in which {\bf specific} ways the technology could fail?).

\subsubsection{Privacy and Performance.}

System design choices might lead to various trade-offs between privacy and performance, as discussed below:

\paragraph{Synchronisation Scalability:}
Distributed robust and trustless systems tend to have poor performance due to the complexity of
achieving synchronisation --- this requires both a logical consensus on what transactions should be included
in a new block to be added
to the blockchain and a technical synchronisation of all of the nodes in the distributed ledger \cite{coulouris,lamport}.
As the number of elements requiring consensus increases,
the time taken for consensus may increase.
Similarly, if complex cryptographic procedures for improved privacy are incorporated into the consensus mechanism
then the time taken for consensus is likely to increase
(and time to detect fraud may increase).

\paragraph{BlockChain Compression:}
\label{subsec:blockchain:compression}
%
One major topic of interest (and also a direct metric of achievement in blockchain privacy)
is that of blockchain {\em compression}.
Many privacy techniques can be seen as a method for compressing (reducing the amount of data stored on)
the blockchain, in that only a small amount of information is stored on the blockchain (visible to all participants)
and an extra, larger, amount of data is stored of communicated off-chain (visible to a smaller subset of participants).
For example in Payment Channels \cite{paymentchannel},
only final digital signatures need to be published on the blockchain, and other data does not need to be published.
In Stealth Address techniques such as are used in CryptoNote \cite{CourtoisMercerStealthICISSP} (see Sections~\ref{MostBasicStealthBasics} and \ref{RingSignaturesCryptoNoteVsMobius}),
an Extended Public Key (EPK) of the recipient is sent privately by the recipient to the sender, 
and this EPK will never appear on the public ledger.
Compression is also a key feature in the design of M\"{o}bius (see Section~\ref{MobiusNotGoodStealthAddress})
and in general compression has important consequences for the overall performance of a DL because the data stored on the blockchain
must be replicated on every node in the DL.

\subsubsection{Privacy, Settlement and Liquidity.}
\label{TokenPrivacyProblem:as:LiquidityEconomy}

A major question in the payment and settlement of crypto currencies is the question of liquidity.
This has two aspects: (i) liquidity-saving mechanisms (LSMs) that reduce the overall amounts
of crypto coins that must be transferred by calculating and settling net rather than gross amounts,
and (ii) the ability to dispose immediately of settled crypto coins which
are guaranteed to be available. Here we consider the latter aspect, and do not have good news:
both of the systems\footnote{And also ZeroCash.} that we study (CryptoNote and M\"obius)
trade privacy for the availability of crypto tokens in the form of finally settled and spendable coins.
For example in M\"obius the recipient's identity is
obfuscated by
``mixing'' together a certain number of identical transfers,
so that it is not possible to know connect the identities of senders and the identities of the recipients.
However, this only works if there is a sufficiently large number of participants willing to participate in
this mixing activity.
A recipient might be required to wait until a sufficiently large number of transactions are deposited
into the same smart contract and ready to be withdrawn.\footnote{A related issue is that a bank may wait
in order to withdraw money from a certain contract
for as long as they don't need this money, but not longer,
potentially leading to
timing side channels that may be exploited to more or less reliably
relate the identities of the participants to other events in financial systems.
}
This is discussed in more detail in Section~\ref{MobiusOptions1}.

\label{sec:TokenCrytoAssets}

\newpage

\section{Crypto Methods for Blockchain Privacy}
\label{sec:cryptomethods}

{\colo{black}

Here we review some key cryptographic methods (``crypto methods'') for blockchain privacy:
{\em digital signatures}, {\em ring signatures},
{\em key management and key derivation techniques},
and {\em stealth addresses} (also known as {\em stealth keys}).
Since not all significant details can be included,
we will focus on aspects that might be misunderstood and require elucidation.
Some important observations to start with (further explained below) are:

\vspace{6pt}

{\em
Observation 1:
Digital signatures 
provide elegant ways to separate the control of assets or ability to spend (using ``private'' keys)
from the ability to verify the financial transactions (using ``public'' keys).

Observation 2:
Blockchain transactions are {\bf not} encrypted by default and are largely intelligible
so that the financial integrity of the shared ledger can be audited
independently by numerous network or blockchain participants.

\vspace{6pt}

Observation 3:
Whereas pseudonymity was the initial layer of obfuscation in blockchain technology
(concealing the actual owners of public keys), ring signatures now add another layer of obfuscation.
We will explain how ring signatures allow the signer
\footnote{
\label{WhosePrivacy}
In CryptoNote the signer will be the sender transferring coins and this mechanism is protecting the privacy of the signer.
In M\"{o}bius on the contrary it benefits the receiver 
who signs a withdrawal from the smart contract.
}
 to remain anonymous in a very strong sense,
 but only within a certain relatively small set of users, which can be eroded with time.
}

\vskip-5pt
\vskip-5pt
\subsection{Digital Signatures}
\label{DigSignatures}
\vskip-3pt

A typical setup for public key cryptography is that a user has a private key $a$
and a public key $A$ that is shared with other participants.
Assuming this user creates a Digital Signature $\sigma$ over a message $m$, this can be described
as a mathematical puzzle or an equation to solve:

$$
Verif(A,\sigma,m)=0
$$

such that only one unique\footnote{We call this property non-repudiation or imputability.}
entity or person (the legitimate signer who knows his or her private key $a$)
is able to solve this puzzle (see Section~\ref{sec:puzzle}).
By contrast, once $\sigma$ is given anyone
who knows the public key $A$ can check
if the solution is correct by checking the verification equation.
For a crypto currency, a digitally signed transaction $m$ produced by the owner of $a$ can be verified and
accepted by anyone,
and coins can be irreversibly transferred from the owner of $a$ to a new recipient $B$
stated in the signed transaction,
which $B$ will typically be another public key.
This is very much like signing a cheque except
that a digital signature is used.
}

\colo{black}
\vskip-5pt
\vskip-5pt
\subsection{Solving the Puzzle and Security Requirements in Digital Signatures}
\label{sec:puzzle}
\label{DigSignaturesReq}
\vskip-3pt

It is important to see that the term ``solve the puzzle'' has more than one meaningful definition.
It is about creating a signature $\sigma$ such that $Verif(A,\sigma,m)=0$.
The function $Verif()$ must be sufficiently complex to resist sophisticated attack, i.e. to ensure that only the owner of the key pair $a, A$ can produce the signature $\sigma$ and this signature cannot be forged by an attacker.
The seminal paper of Goldwasser-Micali-Rivest \cite{DS84} was the first to provide rigorous definitions of different levels of attack (forgery)
and therefore a way to measure the strength of protection provided by different versions of the function $Verif()$.
``Universal forgery'' is where an attacker can create any message of his choice 
and successfully forge a signature $\sigma$  to make it appear that the message comes from the owner of the key pair $a, A$.
In real life crypto currency use cases, the main threat would be a so called ``selective forgery'':
signing some specific messages and irreversibly transferring funds.
In real life crypto currency use cases, the main threat would be such ``selective forgery'' irreversibly transferring funds.
Then we have an ``existential forgery'' an attacker can successfully forge a signature for {\bf any} or at least one message $m$
even though this is then likely to be random or irrelevant.
In cryptography, the philosophy is that this (stronger) security definition is required systematically.
Interestingly, contrary to the claim inside the paper \cite{DS84}, it did not provide the ``most general'' or the strongest possible definition.
This is known as the problem of ``duplicate signatures'' --- producing a duplicate signature for a message that has already been signed by the legitimate signer. Many digital signature schemes including ECDSA used in current blockchains systems \cite{ECDSA} remain vulnerable to this type of attack.

\vskip-5pt
\vskip-5pt
\subsection{Ring Signatures}
\label{RingSignatures}
\vskip-3pt

{
In {\bf ring signatures}, we have a group of $N$ signers\footnote{Who do not need to be present,
and whose identity can be used without their permission.} and using this simplified description the equation can be written as

$$
Verif(A_1,\ldots A_N,\sigma,m)=0
$$

where anyone who knows {\bf just one} of the private keys $a_i$ 
can solve the puzzle to create $\sigma$, and once $\sigma$ is known anyone can check that the solution is correct (thereby proving that one of the
owners of $A_1,\ldots A_N$ is the signer of $m$).

\vskip-5pt
\vskip-5pt
\subsubsection{Ring Signatures: Anonymity vs. Linkability.}
\label{RingSignaturesAno}
\vskip-3pt

Ring signatures are ``signer-ambiguous'' in that there is no way to know which signer signed the message
--- if $N$ is very large, this offers very strong anonymity guarantees, but clearly no anonymity at all if $N=1$.
This is used in some digital currencies, for example in CryptoNote-based coins such as Monero.
The following
{\bf simplistic} example
(which is not yet quite complete or secure, as we explain below)
illustrates one basic and quite standard way (cf. CryptoNote protocol)
in which ring signatures can be used in order to authorize
monetary transfers:

\colo{black}
\begin{itemize}
\item
A user $A_x$ who wishes to transfer coins to user $A_y$ will create a transaction containing
the recipient or destination\footnote{It can be any public key
and does not have to be any of $A_1,\ldots, A_N$ in the ring or group specified above,
and it could furthermore be something different than a key,
for example a script or a smart contract in Ethereum identified by a cryptographic hash.} public key $A_y$,
and sign this transaction using a ring signature for a 
well-chosen\footnote{It makes sense only to include public keys which
currently have a balance at least equal to the monetary value to be transferred.}
subset of $N$ users.
\item
The ring signature proves that the transaction
comes from one of the $N$ users without revealing the precise sender's identity.
\item
All users scan every transaction to see whether it contains a transfer to them,
using their public key only.
\item
Here, the recipient (e.g. characterized by his public key) of a transfer is known and recorded on the blockchain.
The recipient's private key can be used to transfer the coins further.
\end{itemize}

In this example, the ring signature is able to conceal the identity of the sender and not more.
All other data about this transaction are visible.
The sender is only known to be one of a (hopefully large) group of users.
Since the sender can create a ring signature using other user's identities without their permission,
the sender should be able to choose a large group within which to hide.

In general, a ring signature has a limited objective: to ensure the privacy of the signer.
If the signer is the sender, then the receiver's privacy is not ensured, and {\em vice versa}.
%

{\colo{red}
{\bf Crucial Difficulty:}
Moreover there is a {\bf major problem} with the approach above which was not solved or not yet.
a key problem is that during the lifetime of this system\footnote{As in the simple example above, similar to CryptoNote.}
we do not know which coins are already spent or not\footnote{In M\"obius this would be rather like which deposits were already withdrawn.}
The answer is quite surprising, we do not need to know (!),
as long as we have {\bf the ability to prevent every single user from spending his coins twice}.
}
For this we will need ``linkable ring signatures'' further\footnote{Moreover even this may be not sufficient in order to achieve financial integrity
and avoid double spending, which question is studied below in Section \ref{LinkableRingSignaturesDifficulties}.}
 explained below
in Section \ref{LinkableRingSignatures}.

\vskip-5pt
\vskip-5pt
\subsection{Linkable Ring Signatures}
\label{LinkableRingSignatures}
\vskip-3pt

In this type of ring signature,
an additional mechanism is used to prevent double processing
of any UTXOs or deposit. 
The digital signature contains an additional piece of data that assists in the detection of duplicate transactions,
so that subsequent duplicates can be disallowed.
This mechanism degrades the signer privacy very slightly:
just the right amount in order
to detect errors or fraud and not more.
A cryptogram\footnote{This word means that it is generated using a cryptographic technique.
It is not meant to be used to transmit message to be ``decrypted'' by another user but rather to
ensure correct operation and avoid prevent double spending attacks. }
known as {\bf key image} (also known as a ``linking tag'') $I$
is produced which is uniquely linked to the sender's private key $a$,
using
a one-way function $f$ (whose value does not reveal $a$),
a value $d$ that is studied later in Section~\ref{RingSignaturesCryptoNoteVsMobius},
and maybe other information (omitted from this simplified explanation)
so that if $I$ is the same for any two transactions only the first will be processed:

$$
I=f(a,d)
$$
The verification of this key image $I$ is already embedded inside the primary equation {\em Verif} used to verify the signatures.
To show this we can write:

$$
Verif(A_1,\ldots A_N,I,\sigma,m)=0
$$

In practice $I$ is frequently omitted because by convention $I$ will be already included
as a mandatory field inside the specification of $\sigma$.

We need to stress the importance of the fact that $I$ is present and that all checks
which involve $I$ must be performed systematically by all the parties involved.
Omission would be fatal, see Table \ref{Table2} 
page \pageref{Table2}, and would lead to double spending.
This double-spending detection is done by a seemingly very
simple algorithm sometimes called RS.Link \cite{Mobius} or LNK \cite{CryptoNote20}
which essentially detects if a given $I$ was previously used and recorded on the blockchain.
If it has previously been seen, this is a double-processing attempt and the blockchain should
simply reject the second (later) spending attempt.

This apparent simplicity hides a serious security question.
There should be no way for the signer to cheat; for example to produce a fake key image $I$ which would compromise the detection (and rejection)
of a duplicate transaction and therefore enable double-spending of his coins.
This is ensured by the fact that the key image is a part of the (ring) digital signature, and its correctness is checked by verifying the appropriate signature verification equation(s) $Verif$.
Informally, the whole digital signature is also a proof of knowledge, showing that the person spending coins knows the right private key, and that this same key is also used to compute this exact key image.
It is a subtle solution
and it is not easy to see if a linkable ring digital signature scheme does all this correctly
(thereby making this type of fraud impossible).
The only method currently known which gives a comprehensive assurance is a detailed cryptographic security proof.
This is necessary and yet not sufficient.

%

\vskip-5pt
\vskip-5pt
\subsection{Difficulties with Application of Ring Signatures}
\label{LinkableRingSignaturesDifficulties}
\vskip-3pt

In fact even in the presence of such a security proof there may be doubts as to whether it is actually correct,
and whether the abstract security result has been interpreted correctly, and whether it is a meaningful security result
that can be applied correctly in a real-life setting. In general, both CryptoNote and M\"{o}bius are quite complex and intuitively
it is very hard to see if they are actually secure. Another question is whether the system has precisely and correctly specified
in which circumstances $I$ should be identical in order to trigger rejection. In both cases the authors
have modified and adapted a ring signature scheme developed by different cryptography researchers,
which means that their security can no longer be endorsed by the original authors and should not be taken for granted.

{\bf Further Difficulties:}
In general a (linkable) ring signature must be integrated with the rest of the crypto currency implementation
in order to address the issue of financial integrity correctly and all aspects of the design must work together.
It is NOT true that an integrity property such as above is easy to achieve with just some crypto ``magic'' solution
such as a ring signature used as a black box. 
It is NOT sufficient for the ring signature scheme to be cryptographically secure (or provably secure) and
a practical application of a cryptographic solution may invalidate\footnote{The financial integrity
may be broken for reasons of the inadequacy of the cryptographic modelling which is used
in order to argue or prove security. }
 the security result. 
This happens for example when the tool is used in situations not anticipated
\footnote{For example, due to special events which may have been artificially excluded from security modelling by crypto researchers
or considered as happening with a negligible probability.}
by the designers\footnote{For example in CryptoNote 
we would 
require a guaranteed 1:1 correspondence between public keys and unspent transaction outputs.
This is likely to lead to real life events which the cryptographic model could have omitted
to consider.}
and whether we are able to accurately and convincingly
reduce a broader financial integrity of a payment system
to a pure crypto security property (an impossibility result).

\newpage

\vskip-5pt
\vskip-5pt
\subsection{Key Management and Key Derivation}
\label{ReceiverPrivacy}
\vskip-3pt

Key management and key derivation techniques are primarily used to enhance the privacy of the receivers of assets.
The main goal is to achieve a disconnection between
the cryptographic identity (usually in the form of a ``master'' cryptographic public key)
and the ``child'' identities actually used in one or (for better privacy) several transactions
to transfer coins that are recorded on the blockchain.
Key questions to study are (i) blockchain compression (blockchain storage space
is an expensive resource),
(ii) how the ``master'' keys can be communicated using additional communication channels and
how to secure their authenticity  and
(iii) how the receiver is able to see the incoming payments (independently of the ability to spend them)
given that we aim to conceal them as much as possible\footnote{But not more:
we are not willing to risk that the receiver will not see some incoming payments,
which is a potential problem --- see Section~\ref{MostBasicStealthFourMistakesPoint4}. }.


\subsubsection{Key Management Introduction.}

The necessity to use several pseudonyms comes from the poor anonymity of
existing blockchain systems.  Attempted solutions to the anonymity problem can create a
further problem: the fragmentation of assets which are stored at various pseudonymous/public keys
which leads to high fees when transferring those assets and may also lead to ``consolidation'' transactions which
may reverse the privacy obtained earlier.
Practical implementations of the privacy provisions in M\"{o}bius \cite{Mobius}, CryptoNote \cite{CryptoNote20} and ZeroCash \cite{ZeroCash}
all either already have or are likely to have this problem.
We are not yet aware of a truly well-designed anonymity solution for blockchain privacy that would be
devoid of problems and inherently secure\footnote{This would mean secure in every case, independently
on how it is implemented and used. Provable security does not solve this problem.
}.

Advanced key management techniques were first used in Bitcoin
to diversify the keys to be used in different transactions,
and then to develop hierarchical key management methods 
\cite{CombinationAttacks,BitcoinKeyManagementHDWallets}
with so called ``Audit Capabilities''.
These capabilities allow third parties to derive public keys
from certain ``master public keys''  without knowledge of the private keys.
More recently,  ``Stealth Address'' techniques have become popular
and these techniques have an additional major objective:
to increase privacy for the receivers of payments.
We refer to \cite{CourtoisMercerStealthICISSP} and to \cite{CombinationAttacks,CombinationAttacksTexas,BitcoinKeyManagementHDWallets,UsabilityBitcoinKeyMgmnt,GutoskiHDWalletFix}
for a detailed study on the topics of key management emerging in crypto currencies and in the industry at large.
In this paper we focus on the basic features and how key derivation could or should be performed
in order to benefit specific use cases involving asset storage, exchange and settlement.

\subsubsection{Key Management with Audit: Main Principle.}
\label{KeyMgmntAuditPrinciple}

We outline the primary methods used in creating ``Auditable'' key management
(sometimes called ``Type 2 techniques'' \cite{CombinationAttacks,BitcoinKeyManagementHDWallets}).
The property we need is to be able to derive public keys without reference to the private keys, or rather do both in bulk
from a ``Master'' public key\footnote{Sometimes a word ``SEED'' is used instead of the word ``key'',
and sometimes other terms such as ``Extended'' Public/Private Key are used \cite{CombinationAttacks,BitcoinKeyManagementHDWallets}.
In this paper we prefer to use the term of ``Master'' keys.
}
and a ``Master'' private key, together with
two key Child Key Derivation (CKD) functions:
a ``Private CKD'' function and a ``Public CKD'' function
as illustrated in Figure~\ref{BitcoinHD2WalletsPrinciple}.

\begin{figure}[!h]
\centering
\begin{center}
\includegraphics*[width=3.8in,height=1.6in]{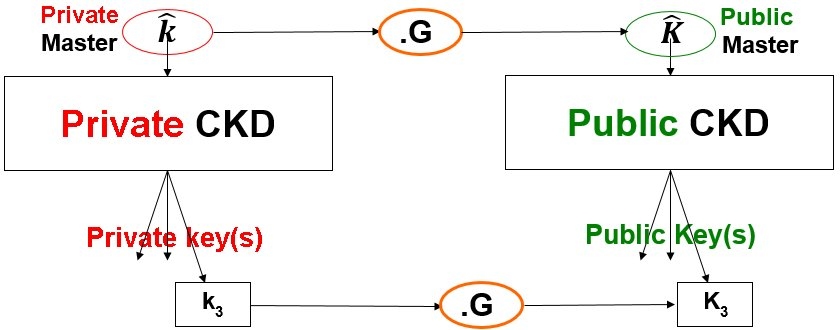}
\end{center}
\parbox{3.8in}{\caption{\label{BitcoinHD2WalletsPrinciple}
General key derivation principle from \cite{CombinationAttacks}
with modified notations.
It must be such that the diagram commutes
and public keys $K_i$ obtained either way are identical.  The generator $G$
produces public keys from private keys, e.g. $K_3 = k_3\cdot G$.
}}
\end{figure}

\vspace{-12pt}
The crucial property is then
that the corresponding lower level keys match.
This property means that our diagram in Figure~\ref{BitcoinHD2WalletsPrinciple} should commute,
i.e. that
$CKD_{public,i}(\hat k \cdot G)  = (CKD_{private,i}(\hat k)) \cdot G$.
In \cite{CombinationAttacks} it is also required that
all the 4 derivations in Figure~\ref{BitcoinHD2WalletsPrinciple} 
are one-way functions and therefore for example the public master key/seed should {\em not} reveal
the private master key/seed (nor indeed any of the private keys).

It possible to see that such solutions
can be deployed at several levels and lead to
Hierarchical Deterministic (HD) solutions where more complex multi-level diagrams will also commute
\cite{CombinationAttacks,BitcoinKeyManagementHDWallets,WuilleBIP32Spec}.
Such systems can be studied in terms of ``security domains'' or partially ordered sets
which allows for information flow analysis \cite{CombinationAttacks}.

\vskip-5pt
\vskip-5pt
\subsubsection{Key Management with Audit: Solutions.}
\label{KeyMgmntAuditPrincipleTwoMethods}
\vskip-3pt

The main idea in constructing such key derivation schemes
is that both derivation functions are essentially the same (and have the same inputs).
Private keys and public keys are modified in a consistent way,
and algebraic homomorphic properties of Elliptic Curve Cryptography (ECC) are used.
There exist two basic methods:  multiplicative and additive.
Their history goes back to June 2011, a date when the forum thread ``Deterministic Wallets''
was started\footnote{Cf. \url{https://bitcointalk.org/index.php?topic=19137.0}} by Greg Maxwell.
The multiplicative solution was proposed first and was used in various systems such as Electrum.
Since April 2013 there was a shift towards the additive method,
which is claimed to be faster and easier to implement \cite{WuilleBIP32Spec}, and which became
standardized inside the BIP032 specification.\footnote{\url{https://github.com/bitcoin/bips/blob/master/bip-0032.mediawiki}}

\begin{defi}[The Multiplicative CKD Method]
\label{TwoKMgmntAddMult1}
The multiplicative method
has a Private CKD function:

\vskip-4pt
\vskip-4pt
$$
k(x)=(\hat{k}\cdot H(\hat{k}.G,x)) ~~\mod Q
$$
\vskip-3pt

and the multiplicative method Public CKD is:
\vskip-4pt
\vskip-4pt
$$
K(x)=H(\hat{K},x).\hat{K} ~~~~\mbox{in~~} E(\F_P)
$$
\vskip-1pt

where $\hat{K}=\hat{k}.G$, and where $H()$ is a hash-to-point function with output reduced $\mod Q$,
and $K(x)$ will be a point on the elliptic curve $E(\F_P)$.
\end{defi}

\begin{defi}[The Additive CKD Method]
\label{TwoKMgmntAddMult2}
The additive method has a Private CKD function:

\vskip-4pt
\vskip-4pt
$$
k(x)=(\hat{k}+H(\hat{k}.G,x)) ~~\mod Q
$$
\vskip-1pt

and the additive method Public CKD is:

\vskip-4pt
\vskip-4pt
$$
K(x)=\hat{K}+H(\hat{K},x).G ~~\mbox{in~~} E(\F_P)
$$
\vskip-2pt
\end{defi}

We refer to \cite{CourtoisMercerStealthICISSP} for a proof of correctness of these two methods.\footnote{M\"obius (\cite{Mobius}, Definition 2.1) uses an
additive method, where a hash is added to the private key.
The paper uses unusual multiplicative notations which are rarely used for ECC and which make it not completely apparent.}

{\bf Security.}
All these methods are subject to ``bad random\footnote{Exploitation of
real-life events due to the imperfect generation of numbers
that are required to be random.}'' and other cryptographic attacks
in which coins may eventually be stolen {\bf if} certain specific events happen \cite{CombinationAttacks}.
They are potentially secure in ideal circumstances where all random numbers are perfectly random,
where keys passwords and identities are not re-used, and where audit keys are not leaked to attackers.
In the real world a certain small percentage of keys will\footnote{See for example \url{http://blog.bettercrypto.com/?p=1099}} be compromised.
Later in 2014-17 two improved solutions with extra robustness against attacks were proposed:
a multi-key multiplicative method of \cite{GutoskiHDWalletFix}
and a mixed additive-multiplicative Robust Stealth Address solution of \cite{CourtoisMercerStealthICISSP}.

\subsubsection{Further Problems with CKD.}
\label{WhatIsWrongCKD}

There are at least two problems with using a simple Child Key Derivation (CKD) technique as illustrated in Figure~\ref{BitcoinHD2WalletsPrinciple}.
First, it uses only {\em symmetric cryptography}, where both CKD functions must share the same secret (such as $\hat{K}$),
which strongly limits what can be achieved
(many advanced security techniques require asymmetric cryptography).
Second, it lacks ``freshness'' for different operations
and operates in an old-fashioned 
uni-directional ``Key Transport'' communication model\footnote{
Key Transport was one of the dominant methods of secure communication used in encrypted Internet communications.
It has been discontinued since 2014 by The Internet Engineering Task Force \cite{DropKeyTransport}.}
and it does not exploit the full power of interaction between the participants which would enable better security\footnote{
Broadly speaking, better security is achieved through limiting the damage which can be done
if some particular devices or keys are compromised
and extends to the resistance against quantum computers by fragmenting the attack,  creating larger numbers of keys,
and reducing the potential economical gain from compromising one key.}.

\colo{OliveGreen}

\subsection{Asymmetric-Symmetric Stealth Address.}
\label{MostBasicStealthBasics}
\label{MostBasicStealthSymmetricVSAsymmetric}

The basic Stealth Address (S.A.) technique
was invented by 
user ``bytecoin'' in the Bitcoin forum on 17 April 2011
where it immediately attracted the attention
of a Bitcoin core developer Mike Hearn \cite{bytecoinStealth}. In summary:

\begin{itemize}
\item
The recipient has a private key $b$ and public key $B=b.G$ and the
sender has a private key $a$ and public key $A=a.G$ 
\item
A Diffie-Hellman key exchange \cite{DH} allows both the sender and the recipient to compute a shared secret value $S$ while
communicating over an insecure communications channel such that:\footnote{And $S \in E(\F_P)$ --- i.e. S is a point on an elliptic curve over a finite field.}

$$
S=a.B=b.A
$$

The sender calculates $S=a.B$ and the receiver calculates $S=b.A$.
\item
Both sender and receiver use the same one-way hash function $H$ and the same modulus $Q$ to hash the shared secret to get an ephemeral private key $c=H(S) \mod Q$, and
to generate an ephemeral public key $C = c.G$ for the subsequent transfer.
\end{itemize}

\vspace{6pt}
Figure~\ref{MostBasicStealthFourMistakes} illustrates a possible variant of how this might operate in practice (with several problems that will be discussed below):

\vspace{6pt}
\begin{description}
\item[Step 1:] The sender retrieves the recipient's static public key $B$ from the blockchain or elsewhere (e.g. the Internet).
\item[Step 2:] The sender computes the ephemeral S.A.  $C = c.G$ (where $c=H(S)$ based on the calculated shared secret $S$) and sends a blockchain transaction to that address.
\item[Step 3:] The sender informs the recipient (via an off-chain channel) that a transaction has been sent [optional].
\item[Step 4:] The recipient calculates the ephemeral S.A. $C = c.G$ (where $c=H(S)$ based on the shared secret $S$) and retrieves the transaction from the blockchain.
\end{description}


We call this technique ``Asymmetric-Symmetric'' 
because it
uses asymmetric cryptography,
yet it is highly symmetric in the way in which both
participants operate; for example both can spend the coins, which is not ideal for the receiver as we will see below. *
%
%


%
%

This sort of technique
or similar simple variants are widely used. 
Now the devil is in the details and we will see below that this technique admits numerous variants which are far from being
equal with respect to usability, security and privacy.
We make the following observations on the original S.A. method (see also \cite{CourtoisMercerStealthICISSP}):

\begin{figure}[!h]
\centering
\begin{center}
\includegraphics*[width=4in,height=2.8in]{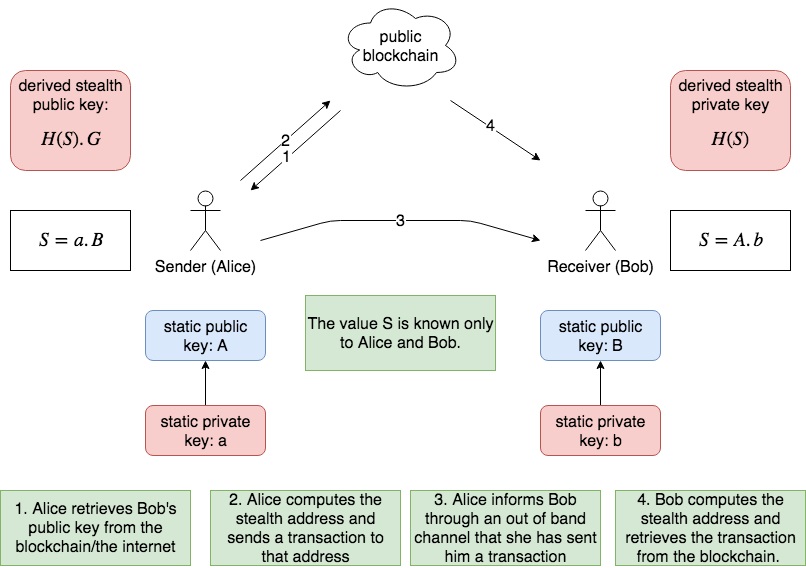}
\end{center}
\parbox{4in}{\caption{\label{MostBasicStealthFourMistakes}
A possible variant of our basic Stealth Address technique which however suffers
from several problems as discussed in the text [picture by Killian Davitt]. 
}}
\end{figure}

\colo{black}
\vspace{6pt}
\begin{description}
\item[Security issue 1: reuse of $A$.]
\label{MostBasicStealthFirstMistake}
The first problem occurs when the sender
uses his or her permanent identity $A$ to receive several payments,
which might become linked to each other\footnote{This may not be obvious because additional data are not made public.}.
Later variants of this technique such as 
\cite{CryptoNote20} and 
\cite{ToddStealth} make it
clear
that we need to do something different.
What is recommended in \cite{CryptoNote20,ToddStealth}
is to use a random ephemeral ``nonce'' key-pair.\footnote{A {\em nonce} is an arbitrary number that can only be used once,
and a {\em nonce key-pair} is an arbitrary key-pair that can only be used once.}
We denote
such an ephemeral nonce key-pair by $r,R$.
In this case the value $R$ must be transmitted with the transaction
which increases the blockchain space required
\cite{MetaDataOpReturn}.\footnote{One method to publish extra data in Bitcoin
is to use the \textsc{op\_return} instruction where
arbitrary data can be placed in the outputs of Bitcoin transactions \cite{MetaDataOpReturn}.}

\item[Security issue 2: ``sender-spending'' attacks.]
\label{MostBasicStealthSecondMistake}

A second serious issue with the original method (previously
discussed in \cite{bytecoinStealth} and fixed in all later
proposals \cite{CryptoNote20,ToddStealth})
is that the coins are stored on the blockchain and can be spent by whoever
knows the private key $c$ --- in this scheme both the sender and the receiver can spend because both can compute the private key $c$
\cite{bytecoinStealth,ToddStealth,CourtoisPrivacyMoneroSlides6}.
Therefore if the receiver does not spend the coins immediately or is offline,
the sender can change his or her mind and take the money back.
This is not always a threat because the blockchain can require additional proof to withdraw the funds
(e.g. in
ZeroCash).

\item[The necessity for out-of-band storage and communication.]
\label{MostBasicStealthFourMistakesPoint3}
In\\
Figure~\ref{MostBasicStealthFourMistakes}, Step 1 illustrates a third issue in using basic S.A. techniques
where
we store the recipient identity on the blockchain.
In fact, Step 1 in Figure~\ref{MostBasicStealthFourMistakes} is often expected to happen off-blockchain
(for example, this is how M\"{o}bius is expected to operate).
For example we visit a merchant's web site and download their
digitally signed permanent S.A. and install the merchant as our recipient (very much like adding a recipient in PGP or
a certificate in a web browser).
Then we can engage in multiple transactions with this recipient without
linking these transactions in any way: {\em neither} to the content of their web site, {\em nor}
to any other payment sent by ourselves or others to the same recipient,
{\em nor} to any other information available in the public blockchain.
%
\item[Recognising incoming transactions.]
\label{MostBasicStealthFourMistakesPoint4}
Step 3 of Figure~\ref{MostBasicStealthFourMistakes} is not necessary,\footnote{This step is advocated yet optional for M\"obius \cite[Sections 3.1 and 8.1]{Mobius}.}
since the recipient can alternatively monitor the blockchain for incoming transactions
(also see \cite[Slides 18--25]{CourtoisPrivacyMoneroSlides6}).
In all cases the receiver can actively monitor the blockchain or other channels for all plausible $A$
and check if somebody is sending coins to some $H(b.A).G$ or in some cases also to some $H'(H(b.A).G)$
each of which he or she can recompute and compare.
If detected, he or she can spend all such transaction outputs (coins).

On the other hand, monitoring the blockchain for the incoming payments means that the private key of the recipient is used more frequently and is therefore
more exposed to theft or side channel attacks, which is not ideal.
The solution is to move towards yet more advanced ``Dual-Address'' S.A. methods,
see Section~\ref{StealthTechniqueDualKey} (which is already implemented in CryptoNote 2.0 \cite{CryptoNote20}).
\end{description}



%


\subsection{Improved Stealth Address Methods}

\subsubsection{Improved basic method.}
\label{BetterBasicStealthTechnique1key}

An improved basic method (which is extended by an additive key management technique
in the sense of Definition~\ref{TwoKMgmntAddMult2}) fixes a couple of the problems studied above.
This method was first described by Nicolas van Saberhagen in the
CryptoNote white paper which appears to be from October 2013 
\cite{CryptoNote20} and in January 2014 it was adapted
to the Bitcoin specification \cite{ToddStealth}.
%

\begin{enumerate}
\item
The recipient has a public key $B=b.G$
\item
The sender uses a one-time nonce pair $r,R$ where $R=r.G$,
and $r$ is 
random mod $Q$. (This avoids disclosing the sender's permanent identity $A$ multiple times).
\item
A Diffie-Hellman exchange allows both 
to compute the same value $c=H(S)$:
\vskip-4pt
\vskip-4pt
$$
c=H(S)=H(r.B)=H(b.R) ~~\mod Q
$$
\vskip-2pt
\item
The ephemeral private key which only the receiver can compute (thus avoiding the ``sender-spends'' attack) is then:
\vskip-4pt
\vskip-4pt
$$
c+b=(H(b.R)+b) ~~\mod Q
$$
\vskip-2pt
and the publicly visible address which will appear on the blockchain
(and which both can compute)
will be no longer $H(S).G$ but $B+H(S).G$:
\vskip-7pt
\vskip-7pt
$$
H(S).G+B=H(r.B).G+B=H(b.R).G+b.G ~~\in E(\F_P)
$$
\vskip-5pt
\item
\colo{OliveGreen}
The receiver monitors the blockchain for
transactions that include 
a publication of some $R$ value
(for example after an \textsc{op\_return} in Bitcoin),\footnote{
The publication of $R$ in a transaction increases memory usage on the blockchain
and could be prone to censorship \cite{MetaDataOpReturn}.
A major alternative
is go back to using some more permanent keys for the sender $a,A$ instead of $r,R$, cf. \cite{CourtoisPrivacyMoneroSlides6}.
cf. \cite{MetaDataOpReturn,CourtoisPrivacyMoneroSlides6}.
}
and for such transactions he or she can compute the private key as $(H(b.R)+b) \mod Q$
and spend the coins.
\end{enumerate}

\vspace{-6pt}
\subsection{Dual-Key Improved Stealth Address Methods}
\label{StealthTechniqueDualKey}

An important enhancement to S.A. methods
are {\em Dual-Key} methods.
The oldest description of such a technique we are aware of
is the CryptoNote paper 
\cite{CryptoNote20}. Figure~\ref{MostBasicStealthTwoMistakesSlide292} illustrates the following short description of a Dual-Key S.A. method
used in many current systems such as Monero (see also \cite[Slides 31--40]{CourtoisPrivacyMoneroSlides6}):

\medskip
\vskip-2pt
\vskip-2pt
\begin{enumerate}
\item
The recipient has a S.A. in the form of two public keys (hence the name ``Dual-Key''):
%
\begin{itemize}
\item
a ``scan'' public key $V$ (sometimes called a ``view'' key  \cite{CourtoisPrivacyMoneroSlides6}), and
\item
a ``spend'' public key $B$.
\end{itemize}
As in the previous notation used above, we have:
$V=v.G$ and $B=b.G$
\item
$V$ and $B$ will be points on an elliptic curve and will typically have 33 bytes,
whereas the scalars $v,b$ will typically require only 32 bytes.


\item
The master stealthy public key advertised by the receiver of coins is $B,V$.
None of these keys ever appears in the blockchain, only the sender and the receiver know $B,V$.
\item
The sender uses a one-time nonce pair $r,R$ such that $R=r.G$, and $r\leftarrow$ random $\mod Q$.
\item
A Diffie-Hellman exchange allows both the sender and the recipient
to compute a shared secret value $c=H(S)$:
$$
c=H(S)=H(r.V)=H(v.R) ~~\mod Q
$$
\item
However the ephemeral private key which only the receiver can compute is:
\vskip-4pt
\vskip-4pt
$$
c+b=(H(v.R)+b) ~~\mod Q
$$
\vskip-3pt
and the publicly visible address which will appear on the blockchain
(which sender, receiver and an auditor --- see below --- can compute) is the address
with public key equal to $H(S).G + B$ and:
\vskip-7pt
\vskip-7pt
$$
H(S).G+B=H(r.V).G+B=H(v.R).G+b.G ~~\in E(\F_P)
$$
\vskip-3pt

\begin{figure}[!h]
\centering
\includegraphics*[width=3.8in,height=1.7in]{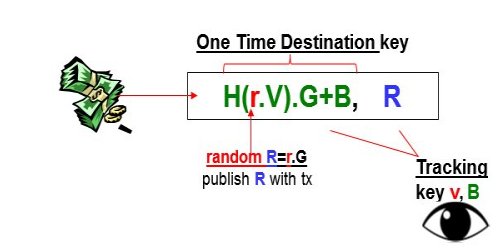}
\parbox{3.4in}{\caption{\label{MostBasicStealthTwoMistakesSlide292}
A fresh destination address in a Dual-Key S.A. \cite[Slides 33--44]{CourtoisPrivacyMoneroSlides6}.
}}
\end{figure}

\item
The auditor, proxy server or read-only (hot) wallet actively monitoring the events,
knows the pair $B,v$.
%
The auditor actively monitors the blockchain for
transactions which include 
a publication of some $R$ value,
and for such transactions he or she can compute

\vskip-4pt
\vskip-4pt
$$
pk=H(v.R).G+B ~~\in E(\F_P)
$$
\vskip-3pt

and see if this $pk$ or its hash appears in the blockchain.

\item
The auditor is not able to spend coins because he does not know $b$.
Only the recipient knows $b$ and can compute $sk=H(v.R)+b \mod Q$ and
spend these coins.
\end{enumerate}

\newpage
\colo{black}

\section{Proposed Solutions Adding Privacy to Crypto Currency}
\label{sec:MobiusvsMonero}

Practical implementations of crypto currency privacy
differ very substantially in how they approach the privacy problem.  For example, consider the following four major directions in blockchain privacy:

\begin{enumerate}
\item
Dash is the simplest low tech co-operative solution which encourages peer mixing.
It does not use advanced cryptography.
 \item
In M\"{o}bius the S.A. technique is extremely basic: money is sent not to a user but to a custodian smart contract,
and ring signatures are used at withdrawal stage providing (with caveats)\footnote{For example, if we
ignore events and S.A. communication which happened before the withdrawal, and also if we ignore the actions of participants
after the withdrawals, for example aggregating different payments into larger amounts.}
very strong anonymity for the receiver.
\item
In CryptoNote 2.0 (e.g. inside the original Monero crypto currency),\footnote{This applies to an earlier
original stage of development of Monero, as it seems that the primary goal of Monero was to implement
the CryptoNote protocol. Monero has later been developed further in order to conceal
the amounts inside the transactions.}
ring signatures are used to conceal the identity of senders,
and a sophisticated dual-key S.A. technique
is used in order to conceal the identity of receivers.
\item
ZCash uses Zero-Knowledge (ZK) proofs.
Until now \footnote{ZeroCash has had a very major upgrade in late 2018.} ZCash used a very simple (not very strong) S.A. technique
which was combined with Zero-Knowledge (ZK) proof techniques for better security.
This ZK proof technology plays the same role as Ring Signatures in Monero,
except that it is stronger: it convinces the blockchain that
some user has the spending private key without revealing which user it is for a substantially larger number of users (say 1 million).
By contrast: in Monero the size of this set would
be restricted to perhaps 4 or 11\footnote{For example the mandatory ring size is now 11 since Monero v8 as of October 2018.} because otherwise the transaction would become excessively long
(and storage on the Monero blockchain is quite expensive).
\end{enumerate}

In this section we consider how cryptographic approaches to privacy have been implemented in practice, with examples drawn primarily from
2 (M\"{o}bius)
and
4 (CryptoNote 2.0).
Both use ring signatures and are based
on traditional widely-studied 1990s cryptography.\footnote{However, security and privacy in a real-life setting is a lot more complex.}
If CryptoNote 2.0 or the current version or M\"{o}bius are insecure, it is very likely that there will be a way to engineer
around the problem and fix it.
In contrast, ZCash is based on cryptographic assumptions which are so unusual and so new,\footnote{
See Section~\ref{sec:cryptomagic}. Furthermore, the assumptions appear far-fetched: they appear to have have been bent in order to achieve a very ambitious result.}
even for professional cryptographers,
that there is a risk that ZCash might cryptographically collapse overnight (and because this technology is so immature, it is likely that it would then be beyond obvious repair).\footnote{
In the same way as most 1990s standardized hash functions, most older stream ciphers based on Linear Feedback Shift Registers (LFSRs),
and most discrete logarithms in finite fields, have been compromised to a vast extent by cryptography researchers.}

\subsection{Implementing Ring Signatures}
\label{RingSignaturesCryptoNoteVsMobius}

CryptoNote and M\"{o}bius operate in different ways: for example, CryptoNote uses a ring signature to
hide the identity of the sender, whereas M\"{o}bius uses a ring signature to hide the identity of the receiver.

\subsubsection{CryptoNote.}
The basic operation of CryptoNote, e.g. in Monero \cite{CryptoNote20}, is illustrated in Figure~\ref{CryptoNotePrinciple1} and explained below:
\begin{itemize}
\item
The intended receiver of a transfer generates a new Stealth Address 
(a method that can generate multiple public/private key pairs) and sends the S.A. 
to the intended sender via an off-chain communication channel.
\item
The sender generates a fresh one-time destination address $DPK_i$ derived from 
the stealth address provided in the previous step. 
\item 
The sender irreversibly sends money (crypto tokens) to this destination address. 
\item
A ring signature is used to hide the identity of the sender by forcibly mixing
the Unspent Transaction Outputs (UTXO) which actually belongs to the sender
with UTXOs 
from some other senders (all of which are visible).
\item
A crucial part of the ring signature scheme used here to authenticate this transaction
is a ``key image'' (also known as a ``linking tag'' --- see Section~\ref{RingSignaturesAno}).
It identifies the sender's private key and it may be used only once in order to prevent double spending,
i.e. multiple use) of any initial UTXOs\footnote{
{\colo{red}
If the system
adheres to certain global constraints on the usage of public keys 
then this should prevent double spending however the exact specification and encoding of what is allowed
is a potential source of ambiguities and implementation mistakes.}}.

\item
Any UTXO might be used as a passive (dummy) input in a several ring signatures before it is finally spent.
 The actual spending of the UTXO might be revealed later to a third-party observer,
 thereby leading to knowledge about previous ring signatures (since the observer can infer that the UTXO must have been passive in previous ring signatures).
 Or the UTXO might be part of a subsequent ring signature, in which case its presence in the previous ring signature can be inferred to  have been passive.  However, an observer might equally well never have such information revealed (no one except that owners would know if certain UTXOs were spent).
 Thus different third-party observers might have different information about UTXOs.  Fortunately the counterparties to a transaction that spends a UTXO will always have full information about that spending; the integrity of the transaction is assured, and there will be settlement finality.
\item  Although a UTXO input to a ring signature appears to come from any one of the users involved in the ring signature, the key image uniquely determines the user for whom further spending of the UTXO is prohibited --- the others are passive (or ``virtual'') and there is no consequent spending constraint on any of the other users involved in the ring. However, there may be unintended consequences for users who are involved as passive inputs into a ring signature, since their involvement is public and all users will known that there is a possibility that each user in the ring signature might really be spending that money at that time.
\item  This mechanism needs to be carefully engineered and carefully implemented: so that the double processing  attempts are always revealed and yet the disclosure of this ``key image'' reveals nothing more than a random number deterministically linked to the identity of the receiver (and that there is no subtle ambiguity such as duplicate signatures that might actually enable double spending).\footnote{For example imagine that the implementation of the $Verif()$ function is not entirely correct in certain rare cases where, perhaps due to a bug in the CPU causing an integer to overflow in a certain way, two transactions with the same $I$ will be accepted. This would enable double spending.}
\item
The receiver's funds are now completely settled.
He can withdraw the amount from the Monero transaction using the receiver's ``private $DPK_i$'' at any later moment.
\end{itemize}

\colo{OliveGreen}

\begin{figure}[!h]
\centering
\begin{center}
\includegraphics*[trim={1pt 15 0 6},clip,width=3.8in,height=1.6in]{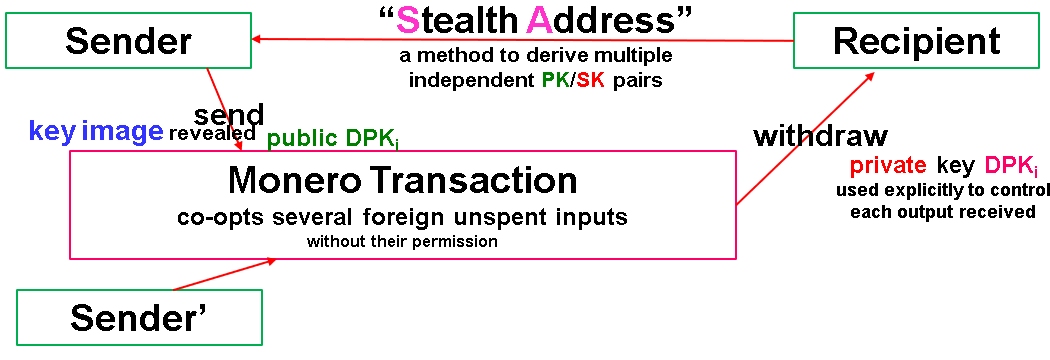}
\end{center}
\parbox{3.8in}{
\caption{\label{CryptoNotePrinciple1}The basic principle of CryptoNote, e.g. in Monero \cite{CryptoNote20}.
The receiver sends a Stealth address to the sender, and Ring Signatures are used at the deposit stage to hide
the identity of the sender. The use of a once-only key image prevents double processing of a deposit.}
}

\end{figure}
\vskip-7pt

{\colo{black}

\subsubsection{M\"{o}bius.}
\label{RingSignaturesMobius}
At the heart of  M\"{o}bius is a deterministic Ethereum contract
which accepts deposits from different participants
and accepts  a ring signature on behalf of each withdrawing user (who remains anonymous).
It also uses the Key Image $I$, 
to protect against processing the same withdrawal twice. 
The basic principle of M\"{o}bius \cite{Mobius} is illustrated in Figure~\ref{MobiusPrinciple1} and is explained below:

\begin{itemize}
\item
Initially it is {\bf crucial} that the sender is able to ascertain the authenticity of some public key which belongs to the receiver,
so that he can receive an authenticated message from the receiver without a risk of that message having been spoofed.
\item
The receiver communicates to the sender his ``master'' public key
in the form of a Stealth Address
transmitted outside of the blockchain.
\item
The sender can now use this master public key in order to compute fresh keys
for which the receiver is able to compute the private keys.
\item
This master key has an important capacity to be used more than once.
For a new transaction, the sender derives (using a child key derivation technique)
a new stealth public key from the receiver's
master public key.
\item
The sender irreversibly sends funds (crypto tokens) to the smart contract (acting as a custodian of funds) using the address ``public $DPK_i$'' denoted by the stealth address derived in the previous step.
\item
The receiver can withdraw the amount from the smart contract via a ring signature using the receiver's ``private $DPK_i$''.  The ring signature
hides the identity of the receiver by co-operatively mixing this transaction with other transactions, and checks that the transaction
does not link to any signature previously used to withdraw from the contract.\footnote{\cite{Mobius} describes two methods, the second of which uses a key image (``linking tag''),
the advantage of the latter method being described in terms of reduced storage cost.}\footnote{It is also stated in \cite[Page 11]{Mobius} that the smart contract checks whether ``the stealth public key is valid (i.e., has an associated secret key that can be used to withdraw from the contract)'' --- it is not at all clear how this could be checked without compromising the privacy of the recipient.}
\end{itemize}
}

\begin{figure}[!h]
\centering
\begin{center}

\includegraphics*[trim={1pt 7 0 7},clip,width=3.8in,height=1.6in]{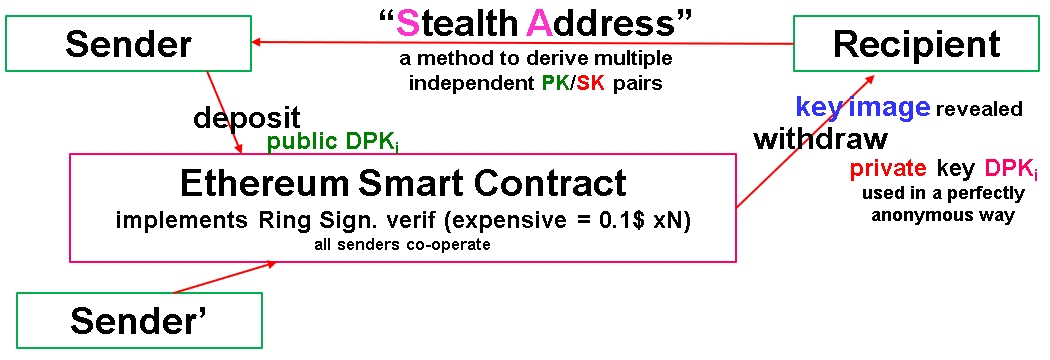}
\end{center}

\parbox{3.8in}{\caption{\label{MobiusPrinciple1}The basic principle of M\"{o}bius \cite{Mobius}.
The receiver sends a Stealth address to the sender, and Ring Signatures are used at the withdrawal stage to hide the identity of the receiver.
The use of a once-only key image prevents double processing of a withdrawal.
}
}
\end{figure}


\subsubsection{Implementing the key image.}
There are other very substantial differences between the type of ring signatures
used in CryptoNote and in M\"{o}bius.
A fundamental difference lies in what is included as $d$ in the key image formula $I=f(a,d)$ (see Section~\ref{RingSignaturesAno}).
CryptoNote implements a {\em modified}
ring signature scheme by Fujisaki and Suzuki \cite{FSRS} where $d$ is essentially empty,
which means that processing the same transaction from the same user twice is always detected.
M\"{o}bius implements another {\em differently modified} ring signature scheme by
Franklin and Zhang \cite{FZRS} where $d$ is essentially the set $R$ of public keys,
which means that processing the same transaction from the same user twice is detected
only if it occurs when processing a withdrawal twice from the same smart contract
with the same group (ring) of public keys.

\vspace{12pt}
{\bf Performance of Ring Signature Implementations.}
The cost in terms of blockchain occupation for most known ring signature implementations grows {\em linearly} with the number of
participants $N$ in a transaction (which also defines the anonymity set).
For example, in M\"{o}bius generally if each smart contract processes roughly one withdrawal for each of the $N$ participants, and if
each  withdrawal has a signature of size $C\cdot N$ , this implies a basic cost of $C\cdot N^2$.
In practice the number of withdrawal events might
be increased due to
the need for additional dummy transactions for obfuscation (a possible practical overall cost might be
say $10\cdot C\cdot N^2$).

\subsection{Implementing Stealth Addresses in M\"{o}bius}
\label{MobiusNotGoodStealthAddress}

In the M\"{o}bius paper \cite{Mobius} the proposed ``Stealth Address'' method has very restricted functionality.
The main intention is said to be to reduce the off-chain communication required to generate shared secrets\footnote{\cite[Section 4.4, Page 11]{Mobius} states:
{\em If communication overhead were not a concern, the system would work the same if, instead of Alice
deriving a new key for Bob in a non-interactive fashion, Bob simply sent a fresh key to Alice to use.}}
and to create multiple freshly-derived public keys based on a master public key, a single shared secret, and
an incrementing nonce.

This is similar in some respects to
an excessively simple\footnote{Compared with advanced S.A. techniques seen elsewhere in the literature \cite{CryptoNote20,CourtoisMercerStealthICISSP,CombinationAttacks,CombinationAttacksTexas,GutoskiHDWalletFix}.
The actual technique described in \cite[Section 2.3]{CourtoisMercerStealthICISSP} is better described as a simple Child Key Derivation (CKD) method --- see Figure~\ref{BitcoinHD2WalletsPrinciple}.}
form of S.A. method proposed in  \cite{CourtoisMercerStealthICISSP}
and interestingly makes some arguments about its security or randomness.
Some highly relevant improvements for M\"{o}bius are suggested in Section~\ref{MobiusPracticeImprove1}
and Section \ref{ImprovedStealthForMobiusandAttackModifR}.

\newpage

{\colo{black}

\section{Discussion}
\label{sec:discussion}

This section provides discussion of issues relating to privacy, liquidity and settlement risk;
issues relating to cryptography in general (especially ``crypto magic''); and issues relating to the implementation of
cryptographic methods.

\subsection{Improving M\"{o}bius}
\label{MobiusPracticeImprove1}

Here we discuss some ways in which M\"{o}bius could be improved, and the problem of providing both
high privacy (optional or on demand) {\em and} high liquidity (also optional),
which is difficult to achieve simultaneously.

\subsubsection{Privacy, Liquidity and Settlement Risk.}
\label{MobiusOptions1}

M\"{o}bius uses a smart contract (a ``tumbler'') to provide privacy via ring signatures.
But the privacy provided by a ring signature is proportional to the number of other available public keys with which
the real transaction's public key can be mixed (if there are no other available public keys, there can be no privacy)
and similarly the privacy provided by a ``tumbler'' smart contract is proportional to the number of transactions
it can ``tumble'' (if there is only one transaction, there is no privacy).  This therefore leads to two important questions:

\vspace{12pt}
\begin{description}
\item[Question 1:]
Do we make privacy obligatory?  i.e. do we enforce users to {\em only} use the
tumbler for all transactions, and do we enforce users to do so {\em frequently} (perhaps including dummy transactions)?
Or is it not obligatory (in which case would there be financial incentives to use the tumbler --- e.g. reduced transactiom fees)?\footnote{
Further examples include: withdrawals could be free after a certain period of time and deposits which are very close to each other in timing
could also be encouraged for better privacy.  This might suggest making both senders and receivers pay some fee as both
can compromise the privacy of others if their actions are excessively contingent to other events in networks and markets.
}

\vspace{12pt}
\item[Question 2:]
If the tumbler smart contract has control of the funds until they are withdrawn, has the tumbler taken legal ownership of the coins as an intermediate step?
Or is the tumbler holding the coins ``on trust'' for the sender (so that if there is a problem the funds will revert to the sender)? Or is the tumbler holding the coins ``on trust'' for the recipient (so that if there is a problem the funds will always be made available to the recipient to withdraw)?
Having the tumbler ``take ownership'' as an intermediate step raises interesting issues since in a truly decentralised system multiple copies of the smart contract might be running one on each node, so the ring signature smart contract might be effectively owned by everyone (and thus everyone might be liable for losses caused by the smart contract malfunctioning). The ability and mechanism to recover coins that were lost or stolen while in the ring signature smart contract is a feature that need further consideration (including how to determine what rectification steps to take under what circumstances).
\end{description}

\vspace{12pt}

\colo{black}

The answer to Question 1 may have important implications for the liquidity of funds (see below), and the answer to Question 2
may have important implications for assessing settlement risk and counterparty risk (see below).\footnote{In the latter case,
counterparty risk may be present for example if the sender becomes bankrupt and the courts determine that any transactions
in the tumbler at the time of bankruptcy were part of the sender's assets (upon which the recipient may have a claim but that claim must
be proved and must be prioritised against other claims).}

Question 2 will no doubt be determined contractually for a permissioned DLT in a corporate environment, but
those corporate users will require highly competent advice from lawyers who understand how the technology works
and what questions to ask; and for crypto currencies handling consumer payments new statutory provisions for consumer
protection may be required.
There is also a technical aspect relating to Question 2: what should the tumbler smart contract do if it detects that there are insufficient
transactions to be tumbled, and it cannot therefore provide a required level of privacy?  Whilst this is not part of the current M\"{o}bius
specification, it should be considered for any implementation of M\"{o}bius.
There are at least four possible options for this technical aspect:

\begin{enumerate}
\item[2.1]
The tumbler continues to process the few transactions that it controls, with little or no privacy benefit, and each recipient may withdraw his or her
funds immediately.   Thus, transactions never fail but privacy is not imposed or not obligatory.
\item[2.2]
The tumbler refuses to process when there are too few transactions on the grounds that it cannot provide sufficient privacy.
Such transactions would fail and the funds would need to be returned to the senders.\footnote{This is potentially easy, since sender UTXOs are not concealed.}
\item[2.3]
The tumbler delays processing the few transactions that it controls until there are sufficient other transactions in the tumbler to guarantee privacy.
This introduces additional and unpredictable delays in liquidity provision to the recipient
(and may also increase settlement risk subject to the issue of ownership,
and perhaps require a backstop of returning the coins to the sender).
\item[2.4]
Each transaction could include a field to indicate how it wishes to be processed in the case of a depleted tumbler --- either to be
processed without privacy, or to be failed, or to be delayed.  The problem with this solution is that both sender and receiver must agree
on the setting for the transaction.
\end{enumerate}

The answer to Question 1 will affect the likelihood of this problem occurring, but cannot avoid the problem completely.
In general these issues are not discussed in the M\"{o}bius paper, which further complicates matters by stating that the set of users in the ring signature should be specified in advance (which is likely to be unacceptably impractical for users).
Furthermore, even the very flexible solution given in option 2.4 above does not solve the problem of minimum privacy standards
(see Question 1 above). Overall, a more precise specification of the operation of M\"{o}bius is required,
following which the ``best'' solution will depend strongly on the requirements of the users
and on how they would relatively prioritise concerns such as privacy, liquidity, and settlement risk.
they depend strongly on the requirements of the users and on how they would relatively {\bf prioritise}
the concerns of privacy, liquidity, and settlement risk.

%
%
%
%
%

\subsection{The Need for Better Stealth Addresses.}
\label{ImprovedStealthForMobiusandAttackModifR}

The main goal of all S.A. methods is to send assets or coins
to a certain ``publicly visible'' master key
in such a way that this key does {\bf not} appear in the blockchain.
The fact that S.A. methods in CryptoNote and even more in the recent 2018 release of ZeroCash
are substantially more complex 
than the simplistic CKD in M\"{o}bius suggests that M\"{o}bius needs something better.

Here we consider a possible attack
 on a modified M\"{o}bius:
\label{CanWeTrustContractToPreserveR}.
M\"{o}bius is intended for use with Ethereum and does not use $R$ yet.
If $R$ is added, and if we want to store $R$ inside the Ethereum blockchain,
we have a problem because by default $R$ is not authenticated.
While we can in general trust the consistency of execution of Ethereum contracts across different nodes,
we cannot trust all individual Ethereum peers to run the same code and behave honestly.
The very first peer in the network receiving funds and the extra data $R$ could cheat,
and modify $R$ making it impossible to ever recover the coins
(unless we implemented some extra deposit recovery --- see Section~\ref{MobiusOptions1}.).
We see three major methods to solve this problem which
might lead to a necessity to augment the M\"{o}bius smart contract system:

\vskip1pt
\vskip1pt
\begin{enumerate}
\item
The sender first publishes $R$ in the Ethereum blockchain.
Then he will wait for several blocks to be mined in order to be assured that this information will not be lost,
and only then send money to the contract (which transaction is irreversible).
\item
A more complex solution is to make the payment dependent on this $R$ and conditional upon
the publication of $R$ in an earlier block.
It is clear that such a mechanism can be implemented in programmable crypto currency
but it is not clear if this works in the current version of Ethereum.
\item
We check that the source code of  M\"{o}bius and Ethereum guarantees the following property:
if the network accepts a signed transaction with $R$, the network nodes must know the actual value of $R$.
This could be difficult when this comes as an upgrade and was not required beforehand.
\end{enumerate}
\vskip-3pt
At the very least we need to carefully audit the code, or modify both the M\"{o}bius protocol and source code.
%
%
\label{MobiusandStealthTechniqueDualKey}
It would be interesting to explore
how a combination of M\"{o}bius and an improved
Dual-Key S.A technique would function.
However, we are not yet ready to
study this question because have a major strategic choice to make
(noting that these choices will have very different effects if they are applied to Monero, Bitcoin,
or a crypto currency using M\"{o}bius.):

\begin{enumerate}
\item
Do we accept that the current M\"{o}bius
with a simple CKD method (see Section~\ref{MobiusNotGoodStealthAddress})
may perhaps be ``good enough'' (see Section~\ref{WhatIsWrongCKD})?
\item
Do we allow the sender to use ephemeral random $R$,
where the sender will send an extra piece of data $R$ with his or her transaction --- currently a popular method in crypto currencies,
e.g. Monero and Section~\ref{StealthTechniqueDualKey}, but vulnerable to the attack mentioned above? or
\item
Do we go back to Section~
\ref{MostBasicStealthBasics}
and consider that it is better to privilege S.A. with more deterministic or
more permanent sender identities which do not need to publish extra information $R$
necessary for the recipient to compute his private key?
This can be done with our standard Dual-key technique
and we would need to replace $r,R$  with some more permanent pair $a,A$.

\end{enumerate}



\label{MobiusandStealthTechniqueDualKeyWithA}

In the light of the attack described above
and the possible necessity to increase the complexity
of the M\"{o}bius smart contract, we tentatively advocate the third solution with $a,A$
which should be an identity previously used by the sender on the blockchain
and which has been published prior to sending money into the contract by the sender.

%
%
%
%
\subsection{Crypto Maturity and Complexity: Crypto Magic}
\label{sec:cryptomagic}

``Crypto magic'' is a folklore term (e.g. \cite{cryptomagic}) which is sometimes used
to denote advanced cryptographic techniques that are perhaps not fully
understood and are sometimes used in order to achieve a very ambitious objective.
\footnote{Such as a large anonymity set in ZeroCash. 
Too good to be true: ZeroCash technology is so complex and advanced that for a long time the system was insecure 
-- it was possible to create money out of thin air -- and nobody has noticed, cf. \cite{ZCashBroken}.}
The research community has not spent substantial time and effort trying 
to evaluate the security of these solutions, 
nor have professional cryptographers claimed these solutions to be secure.
This is a substantial adoption barrier (and for good reason!).
In \cite{bodacious} we read: {\em newcomers to the field would logically expect that
the problems that are used in security proofs come from a small set of extensively studied, natural problems.
But they are in for an unpleasant surprise.}.
Although \cite{bodacious} does not apply directly to current blockchain systems, 
a more relevant example is \cite{ZCashBroken}, 
it does present the problem that the modern cryptographic community does not always try to evaluate the validity of assumptions; 
in essence \cite{bodacious} argues that as long as something is ``proven'' secure it is published even if the underlying security assumptions 
are extremely far-fetched and no one really believes them. 

By comparison, traditional 1990s cryptography such as standardized digital signatures and some more complex derived constructions
(such as Ring Signatures and Stealth Address techniques) are better understood and would not normally be classed as ``crypto magic''.
Where there is additional complexity in a derived construction, the schemes in question require
a security proof (e.g. see the end of Section \ref{RingSignaturesAno}).

Table~\ref{tbl:issues} gives advice in the form of ``six general rules'' for implementors of cryptographic systems; these are particularly relevant
for the implementation of M\"{o}bius and are explained in more detail below:

\begin{table}[h]
  \centering
  \parbox{9.5cm}{\caption{\label{tbl:issues}{\em Beware crypto!}  {\bf Six general rules} for cryptographic systems.}  \label{Table1}}
  \begin{tabular}{lp{9cm}}
  \hline
  \\
  $1.$ &Beware ``common security standards'', which may not be safe.\\
  \\
  $2.$ &Beware unreported trivial vulnerabilities.\\
  \\
  $3.$ &Beware cryptographic solutions built on open-source software.\\
  \\
  $4.$ &Beware schemes that do not have a strong security proof.\\
  \\
  $5.$ &Beware schemes with unrealistic and impractical security proofs.\\
  \\
  $6.$ &Beware that security proofs are necessary but not sufficient.\\[0.3cm]
  \hline
  \end{tabular}
\end{table}

\vspace{6pt}

\begin{enumerate}
  \item
 There is a broad problem with cryptography-related security standards being set by people
 who are not  active researchers in cryptography and who are therefore not aware of the
 many vulnerabilities of cryptographic systems.

  As an example, some security ``standards'' inherited from Bitcoin, such as SHA-256 and {\em secp256k1}, are {\bf not safe to use}.

  The official bitcoin wiki claims that
  ``Bitcoin has a sound basis in well understood cryptography''
  \cite{MythsBitcoinWikiUnprovenCryptoDenial}.  Yet this is incorrect.
  Despite the widespread adoption in bitcoin and blockchain systems of SHA-256 and {\em secp256k1}, they cannot be said to be ``sound'' and ``well understood''.

  \begin{itemize}
  \item
   Most hash functions similar to SHA-256 have already been broken by the academic research community, and since SHA-256 is less studied it is less well understood and it cannot be assumed to be
   more secure \cite{MiningUnreasonable}.
  \item
  The specific parameterisation ({\em secp256k1})
  that Bitcoin uses for its elliptic-curve cryptography is not safe: not in the limited sense of safety in respect to the Elliptic Curve Discrete Logarithm Problem (ECDLP)
  and especially not in the general case of being safe for Elliptic Curve Cryptography (ECC).  There is substantial ongoing discussion of
  the safety of {\em secp256k1}, which points to confusion and lack of commonly agreed understanding.  Bernstein and Lange \cite{BernsteinLangeSafeCurves}
  provide an excellent and accessible summary of different approaches to ECC and ECDLP and their conclusion for {\em secp256k1} is that it is not safe.

  \end{itemize}

There are of course commercial imperatives for asserting that cryptographic security standards are ``safe'', and the growing digital economy needs people and firms to believe that cryptography is safe.
However, caution is advised!

  \item
  Many security solutions are flawed, and the cryptography community is good at publishing papers on novel or highly non-trivial attacks,
  but is not so good at publishing trivial problems; either not at all, or not until a long time after the attack or problem was first known.
  Lenstra et al \cite{LenstraRonWasWrong} is an example of an important disclosure of a trivial vulnerability a long time after it was initially known.

 \item
  Where cryptographic systems are developed using open-source software, undetected vulnerabilities in one person's software may
 be assimilated into other software (we might say that open-source development can be ``infected'' by the virus of a cryptographic vulnerability).
 The resulting systems are often asserted and/or assumed to be secure without any proof or support for that assumption.

  \item Many schemes do not have any security proof whatsoever, or may only have a weak security proof.  For example, \cite{ECDSAnoproof} explains that
  the Elliptic Curve Digital Signature Algorithm (ECDSA) \cite{ECDSA}
  which is in widespread use in current blockchain systems, only has one weak security proof.
  \item Many security proofs are unrealistic and impractical, either because they are framed in a highly theoretical context or because their
  assumptions are unrealistic and/or impractical.  Examples include:
  \begin{itemize}
  \item
  Many security proofs are done in a theoretical setting where a problem is said to be reducible to another problem for which no polynomial-time solution
  is known to exist.  Theoretical proofs of this sort do not consider actual security of concrete instances (where multiple practical vulnerabilities may exist).
  \item
 Many security proofs use a random oracle or ideal cipher or similar model, thus they do not take into account pre-computation and other attacks.
 \item
 Some security proofs are simple ``sketches'' that do not specify the assumptions used (and perhaps do not specify the exact result claimed). One example of
 this approach can be found in the Appendix of \cite{CryptoNote20}.
  \item Many security proofs such as in \cite{FZRS} rely on decisional assumptions which can be weaker than computational assumptions.
  \item Many security proofs use rewinding techniques and suffer from the so called $\varepsilon^2$ syndrome, which means that a factor
  of $\varepsilon^2$ degrades the security reduction. As a result, the system can be simultaneously theoretically secure and practically
  insecure (i.e. broken).
  \end{itemize}
  \item
  Even in the best case (that security proofs such as found and studied in \cite{Mobius,CryptoNote20,FSRS,FZRS}
are tight and have a realistic/concrete setting) it is possible to see
that security proofs still do not mean much beyond the fact that a certain digital signature verification formula\footnote{For example, {\em Verif} in Section~\ref{RingSignatures}
of this paper.} is not completely and egregiously incorrect 
(such as allowing one to cheat easily by modifying the Key Image).
Security proofs are necessary and important but they are not sufficient ---
they do not protect against strong random attacks, side channel attacks, pre-computation attacks,
attacks on the privacy of different actors, and implementation bugs (see Sections~\ref{MostBasicStealthSymmetricVSAsymmetric} and \ref{CryptoImplementationPitfalls})
and thus in general do not replace further detailed security engineering work.
\end{enumerate}

\subsection{Cryptography Implementation Pitfalls}
\label{CryptoImplementationPitfalls}

One of the problems with advanced cryptography solutions is that sometimes their security collapses very very badly
if the user omits to check just one small thing (for example if a number is a prime
or if key image $I$ was correctly formatted).
This will require a comprehensive code security audit, and finding bugs in source code is a task that Rice's theorem \cite{rice} tells us
is impossible to automate in the general case.\footnote{In addition to Rice's theoretic result regarding an automatic procedure,
it may be extremely difficult or impossible to
achieve manually, due to the mathematical hardness of deciding whether a computer program has a certain property,
if cryptography is correctly understood, or from incertitude about how the code will be compiled and optimized,
and how it will be executed by hardware.}
Therefore it is interesting  to list implementation pitfalls explicitly.
At least four major issues
(for many of which we had put into light a number of conflicting requirements)
were already discussed in Section~\ref{MostBasicStealthSymmetricVSAsymmetric}.
However there is a lot more.
Table~\ref{tbl:issues2} gives initial advice in the form of ``six implementation pitfalls'' which could befall an implementation of M\"{o}bius or CryptoNote or any other crypto currency seeking to implement cryptographic solutions for privacy, and these pitfalls (which are illustrative and not an exhaustive list) are explained in more detail below:

{\colo{red}
\begin{table}[h]
  \centering
  \parbox{9.5cm}{\caption{\label{tbl:issues2}{\em Beware crypto!}  {\bf Six implementation pitfalls} for cryptographic systems.}  \label{Table2}}
  \begin{tabular}{lp{9cm}}
  \hline
  \\
  $1.$ &Beware failure to digitally sign key parameters. Verify primality and all ECC parameters.\\
  \\
  $2.$ &Beware missing validation or verification steps.\\
  \\
  $3.$ &Beware perturbation attacks.\\
  \\
  $4.$ &Beware the use of common crypto libraries.\\
  \\
  $5.$ &Beware using the wrong verification ordering.\\
  \\
  $6.$ &Beware ``rogue'' third-party code. \\[0.3cm]
  \hline
  \end{tabular}
\end{table}
}

\vspace{6pt}

\begin{enumerate}
\item
If the elliptic curve parameters such as $P$, $Q$, the base point $G$, and the equation parameters $a,b,c,$ and $d$
are not authentic then security collapses and the system can be hacked.  These key parameters should therefore
be digitally signed, preferably using standard Public Key Infrastructure methods such as a static (persistent) signature inside a certificate.
The primality of $P$ and $Q$ should also be verified.
It should be clear who generated these parameters, how they were generated, in order to avoid potential backdoors.
\item
  In ECDSA signatures there is the following implementation danger \cite{NSAandBitcoinECC_Brown_Denial}:
  \begin{itemize}
  \item
  The ECDSA verification of a signature $(r,s)$ includes a check that $r$ is not zero.
  \item
  If this check is dropped, then there is a possibility that party who chose $G$ can have chosen $G$ in such
  a way to make some signature $(0,s)$ valid for a particular message m.
  \end{itemize}
\item
  The same elliptic curve parameters $P$, $Q$,  and $G$
  should be stored in a redundant way with on-the-fly correctness verification
  in order to prevent perturbation attacks
  such as DFA (Differential Fault Analysis \cite{DFAalg}) or RowHammer \cite{RowHammCourt17}.
\item
  In M\"{o}bius, the key image depend on the ring description $R$ which is a set of public keys (see Section \ref{RingSignatures}).
  A subtle implementation pitfalls lie here.
  Imagine that the same crypto library is used to implement both systems leading to certain things being done
  internally in a way not apparent to the developer. 
  In a well-known annotated version of the CryptoNote white paper,\footnote{\url{https://downloads.getmonero.org/whitepaper_annotated.pdf}} at the
  top of Page 17 (``Comments on Page 9'') it is recommended to permute
  public keys using a pseudorandom number generator in order
  not to append our key as the last in order (revealing the sender). 
  Now if we import this seemingly innocent advice to M\"{o}bius, which uses a {\bf different} formula for computing
  that key image which hashes the ring description, we will enable multiple withdrawals
  because uniqueness of $R$ as a string of bytes for the same ring (as a mathematical set of public keys) will be broken.
\item
  There may be a specific order in which different verifications need to be done:
  On Page 10 of \cite{CryptoNote20} it is explicitly explained that the linkability algorithm is run
  {\em after} checking the signature verification equation.
  Is this recommendation important? We are not quite sure. The reason for this could be that
  linking tags need to be stored and checking them first would allow one
  to have a timing side channel in order to detect if some linking tags were perhaps lost
  or not detected as being repeated (the first check is harder to pass).
  Possibly a RowHammer \cite{RowHammCourt17} or other active side-channel attack could perturbate the storage or key images previously used.
  Then a successful attack would need to test which previously-used tags were erased or modified.
  This is a strong attack in the sense that modifying random memory cells at random can allow monetary gain.
 \item
The use of third-party software may introduce unwitting or malicious errors.  For example,
a third-party process (e.g. a {\em rogue M\"{o}bius wallet}) might contain
a virus or malware, which could for example be enabled after installation as a result of a
``social engineering'' attack. 
Thus, after the user has signed a transaction,
the malware could cause that transaction to be altered such as altering the recipient of coins.
A thorough code audit is required for all software, including all third-party software
and runtime anomaly detection and software integrity checks could also be needed.
\end{enumerate}

}

\newpage

\colo{black}

\section{Conclusion}
\label{sec:conclusion}

Distributed Ledger (DL) systems are proposed as a way to engineer future financial systems.
In this paper we have looked at the issue of adding privacy to DL with a focus on how
``Ring Signatures'' and 
related techniques improve the privacy of token transfers. 
We observe that two eminent exemplars (M\"{o}bius and CryptoNote)
use these cryptographic techniques in radically different ways.
Of the two, M\"{o}bius is more recent and operates in a more co-operative way (with permission).
Our initial analysis and discussion of privacy requirements in the context of financial markets
suggests that adding privacy is (somewhat surprisingly) implied by the high price of disclosure (blockchain space)
and/or the 
threat of data mining (for profit).
We have investigated whether these techniques are fit for purpose,
which has led us to an elucidation of privacy requirements for financial and payment use cases.
Substantial difficulties remain as the cryptographic protocols avoid the difficult questions
of (i) setup and trust and (ii) off-chain channels, and furthermore
(iii) it is not fully specified how various systems should operate in practice.
Either way 
it is clear that privacy conflicts and interacts here with the important issues of liquidity and performance
and that speed and privacy are rather extremely difficult to achieve simultaneously.

Cryptographic technologies such as M\"{o}bius and CryptoNote offer the possibility to
improve privacy in trading and settlement environments where trust is low.
Privacy engineering via Ring Signatures is a form of signer obfuscation, but cryptographic obfuscation does not provide the security guarantees
as good as with encryption (cryptography does not solve all practical problems equally well).
Ring Signature obfuscation provides ambiguity whose strength is limited by the number of participants in the ring and can be eroded
with time as more information is revealed (since privacy relating to users tends to be partial and less easy to control).
Overall, crypto currency systems have complex inter-connections between integrity, performance, privacy and functionality; this makes them brittle.
Furthermore, cryptography itself, and its implementation, is fragile: most cryptography results are provisional rather than finally and convincingly secure. The more complex the cryptography becomes, the more difficult it becomes to know if the source code is correct. Although the security of Digital Signatures is well understood, the security of Ring Signatures as modified or adapted in practice is hard to ascertain beyond any reasonable doubt.
Cryptography and security can fail in many ways due to incorrect trust or setup assumptions, specification or coding ambiguities, failures with adverse events during operation, or failures to perform checks. We recommend a full code audit and yet we must warn that it may fail to address all crypto and security problems at an early stage. Clearly better crypto education is required, and also robust security engineering under the principle that various assumptions can and probably will fail.  Our advice is best summarised in Section~\ref{sec:discussion} as ``Beware Crypto!''.

\newpage

\colo{black}

\vfill
\pagebreak




\end{document}